\documentclass[pdflatex,sn-nature]{sn-jnl}

\date{\today}
\usepackage{amsmath,amssymb,amsfonts}
\usepackage{cases}
\usepackage{subfigure}
\usepackage{graphicx}
\usepackage{dcolumn}
\usepackage{physics}
\usepackage{bm}
\usepackage[section]{placeins}
\usepackage{float}
\usepackage{mathtools}
\usepackage{dblfloatfix}
\usepackage{color}
\usepackage{comment}
\usepackage{bbm}
\usepackage[normalem]{ulem}
\usepackage{amsthm}%
\usepackage{mathrsfs}%
\usepackage[title]{appendix}%
\usepackage{xcolor}%
\usepackage{textcomp}%
\usepackage{setspace}%
\usepackage{multirow}%
\usepackage{manyfoot}%
\usepackage{booktabs}%
\usepackage{algorithm}%
\usepackage{algorithmicx}%
\usepackage{algpseudocode}%
\usepackage{listings}%
\usepackage[font={normalsize,doublespacing},labelfont=bf]{caption}

\usepackage{geometry}
\usepackage{xr}

\newcommand{\muvec} {\hat{\vec{\mu}}}

\begin{document}

\setlength{\parindent}{2ex}

\title[Article Title]{Quantum control of Hubbard excitons}

\author*[1,2]{\fnm{Denitsa R.} \sur{Baykusheva}}\email{denitsa.baykusheva@ista.ac.at}
\equalcont{These authors contributed equally to this work.}
\author[3]{\fnm{Deven} \sur{Carmichael}}
\equalcont{These authors contributed equally to this work.}
\author[4,5]{\fnm{Clara} \sur{S. Weber}}
\author[6]{\fnm{I-Te} \sur{Lu}}
\author[1]{\fnm{Filippo} \sur{Glerean}}
\author[1]{\fnm{Tepie} \sur{Meng}}
\author[1]{\fnm{Pedro B. M.} \sur{De Oliveira}}
\author[7]{\fnm{Christopher C.} \sur{Homes}}
\author[7]{\fnm{Igor A.} \sur{Zaliznyak}}
\author[7]{\fnm{G. D.} \sur{Gu}}
\author[7]{\fnm{Mark P. M.} \sur{Dean}}
\author[6,8]{\fnm{Angel} \sur{Rubio}}
\author[4,5,6]{\fnm{Dante} \sur{M. Kennes}}
\author*[3]{\fnm{Martin} \sur{Claassen}}\email{claassen@sas.upenn.edu}
\author*[1]{\fnm{Matteo} \sur{Mitrano}}\email{mmitrano@fas.harvard.edu}

\affil[1]{\orgdiv{Department of Physics}, \orgname{Harvard University}, \orgaddress{\city{Cambridge}, \state{MA}, \country{USA}}}

\affil[2]{\orgname{Institute of Science and Technology Austria}, \orgaddress{\city{Klosterneuburg}, \country{Austria}}}

\affil[3]{\orgdiv{Department of Physics and Astronomy}, \orgname{University of Pennsylvania}, \orgaddress{\city{Philadelphia}, \state{PA}, \country{USA}}}

\affil[4]{\orgdiv{Institut f\"ur Theorie der Statistischen Physik}, \orgname{RWTH Aachen University}, \orgaddress{\city{Aachen}, \country{Germany}}}

\affil[5]{\orgname{JARA-Fundamentals of Future Information Technology}, \orgaddress{\city{Aachen}, \country{Germany}}}

\affil[6]{\orgname{Max Planck Institute for the Structure and Dynamics of Matter, Center for Free-Electron Laser Science (CFEL)}, \orgaddress{\city{Hamburg}, \country{Germany}}}

\affil[7]{\orgdiv{Department of Condensed Matter Physics and Materials Science}, \orgname{Brookhaven National Laboratory}, \orgaddress{\city{Upton}, \state{NY}, \country{USA}}}

\affil[8]{\orgdiv{Initiative for Computational Catalysis (ICC) and Center for Computational Quantum Physics (CCQ)}, \orgname{Simons Foundation Flatiron Institute}, \orgaddress{\city{New York}, \state{NY}, \country{USA}}}

\date{\today}

\maketitle

\vspace{-8mm}

\textbf{Quantum control of the many-body wavefunction is a central challenge in quantum materials research, as it could yield a precise control knob to manipulate emergent phenomena. Floquet engineering, the coherent dressing of quantum states with periodic non-resonant optical fields, has become an important strategy for quantum control. Most applications to solid-state systems have targeted weakly interacting or single-ion states, leaving the manipulation of many-body wavefunctions largely unexplored. Here, we use Floquet engineering to achieve quantum control of a strongly correlated Hubbard exciton in the one-dimensional Mott insulator Sr$_2$CuO$_3$. 
A nonresonant midinfrared optical field coherently dresses the exciton wavefunction, driving its rotation between bright and dark states. We use resonant third-harmonic generation to quantify ultrafast $\pi/2$ rotations on the Bloch sphere spanned by these exciton states. Our work advances the quest towards programmable control of correlated states and exciton-based quantum sensing.}


\vspace{2mm}

\maketitle

\baselineskip24pt 

Optically-dressed electronic states in quantum materials exhibit remarkable emergent properties and functionalities \cite{Oka2019Floquet,delatorre2021nonthermal}. Intense, nonresonant optical fields have led to the observation of Floquet–Bloch states in topological insulators \cite{wang2013observation,Ito2023buildup}, graphene \cite{Choi2025observation,Merboldt2025observation}, and van der Waals semiconductors \cite{Aeschlimann2021survival,Zhou2023pseudospin,Bielinski2025Floquet}, and of Floquet-engineered phenomena in a range of solid-state platforms~\cite{Sie2015valley,Sie2017large,mciver2020light,shan2021giant,Kobayashi2023floquet,zhang2024light}. To date, experimental efforts have largely focused on the coherent dressing of single-particle band structures of weakly correlated electron gases or the manipulation of local single-ion states.

Coherent control of strongly correlated systems represents the next frontier in the Floquet engineering of quantum materials \cite{delatorre2021nonthermal,Bloch2022strongly}. Theoretical proposals suggest that time-periodic driving can selectively activate or renormalize Hamiltonian terms \cite{mentink2015ultrafast,Valmispild2020dynamically} and stabilize emergent quantum phases, including chiral spin liquids, fractional quantum Hall states, and $\eta$-paired superconductivity \cite{claassen2017dynamical,Lee2018Floquet,Peronaci2020enhancement}. However, the existence of Floquet-engineered states in strongly interacting systems featuring dissipation and decoherence remains an open question. Addressing this challenge requires identifying correlated materials that enable both their generation and detection.

Hubbard excitons in Mott insulators provide a compelling platform for Floquet engineering of many-body states. Unlike conventional excitons, these bound holon–doublon pairs are stabilized by strong Coulomb interactions \cite{Clarke1993particle,Essler2001excitons}. In one-dimensional compounds, they form nearly degenerate even- and odd-parity states, constituting a natural two-level system with distinctive linear and nonlinear optical responses \cite{Kishida2000gigantic,ono2004linear}. Their few-level structure grants direct access to the coherent evolution of the many-body wavefunction, a key feature of Floquet engineering that is typically hard to disentangle from energy shifts, band renormalization, and symmetry-breaking phenomena in more complex systems. Floquet driving on timescales shorter than the exciton lifetime enables deterministic and coherent rotations of the many-body wavefunction within its parity basis, realizing a paradigmatic instance of quantum control.

Here, we demonstrate quantum control of Hubbard excitons in a one-dimensional Mott insulator, Sr$_2$CuO$_3$. Ultrafast, nonresonant midinfrared pulses are used to dress the excitonic wavefunction, generating coherent rotations between two nearly-degenerate states of opposite parity. We probe the evolving quantum state via resonant third-harmonic generation (THG), which is suppressed upon rotation into the dark state, yielding a dynamical renormalization of the harmonic response. Scanning the THG resonance furthermore reveals the presence of Floquet sidebands, a hallmark of coherent periodic driving. The midinfrared pump field induces uniaxial rotations by arbitrary angles up to $\pi/2$, consistent with all-optical $U(1)$ control of the many-body wavefunction. By establishing ultrafast quantum-state rotations in a strongly correlated solid-state system, our work opens new possibilities for quantum control of solids with programmable pulse sequences and exciton-based quantum sensing.

\section*{Nonlinear response of the Hubbard excitons}
\begin{figure}[H]
    \centering
    \includegraphics[width = 0.8\textwidth]{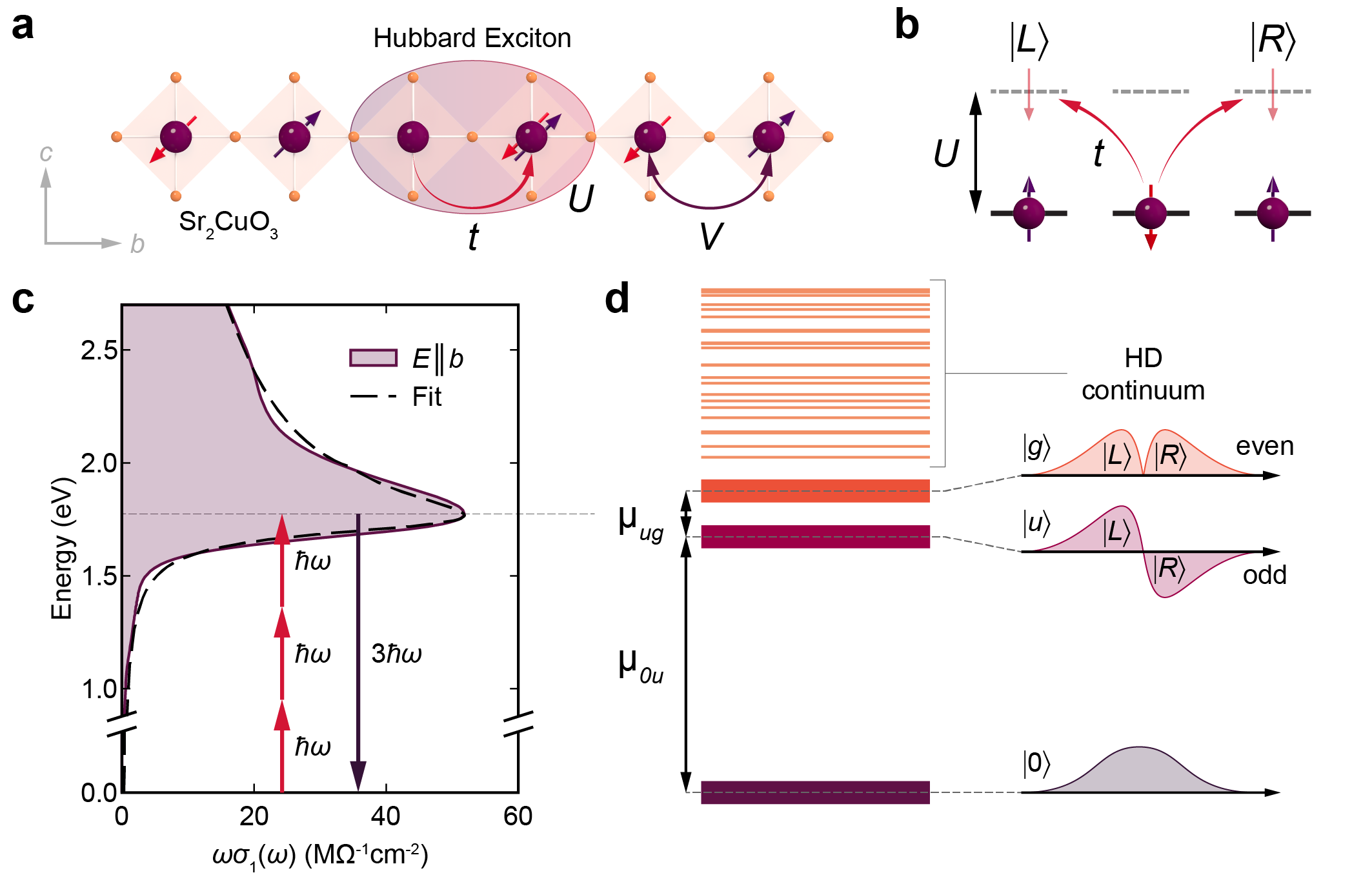}
    \caption{\label{fig:fig1} \textbf{Hubbard exciton and optical nonlinearity of Sr$_2$CuO$_3$.} \textbf{a.} Electrons in the half-filled Mott insulator Sr$_2$CuO$_3$ interact via onsite ($U$) and intersite ($V$) Coulomb repulsion and hop with amplitude $t$. Empty (holon) and doubly-occupied (doublon) sites form bound states, known as Hubbard excitons. \textbf{b.} On-chain electrons can hop left or right into degenerate excitonic basis states $|L\rangle$ and $|R\rangle$, resulting in nearly-degenerate even- ($|g\rangle$) and odd-parity ($|u\rangle$) excitonic wavefunctions. \textbf{c.} Reduced on-chain optical conductivity $\omega\sigma_1(\omega)$ of Sr$_2$CuO$_3$ at 295~K. The Hubbard exciton exhibits strong third-harmonic generation at resonance with the excitonic levels. \textbf{d.} The third harmonic emission involves transitions between ground ($|0\rangle$), odd ($|u\rangle$), and even ($|g\rangle$) excitonic states located below a unbound holon-doublon (HD) continuum. $\mu_{ug}$ and $\mu_{0u}$ are the transition dipoles between odd- and even- exciton states and between ground and odd excitons, respectively.}    
\end{figure}

Sr$_2$CuO$_3$ is a half-filled Mott insulator composed of chains of corner-sharing $\mathrm{CuO_4}$ units aligned along the $b$ crystalline axis \cite{Motoyama1996magnetic} (see \textit{Methods} for sample preparation and characterization). Despite a strong on-chain exchange coupling of $J=2800$~K, three-dimensional magnetic ordering occurs only below $T_N\approx5$~K, indicating extremely weak interchain coupling \cite{Walters2009effect}. The one-dimensional character of the chains is strong enough to induce complete spin-charge and spin-orbital separation \cite{Walters2009effect,Schlappa2012spin}. The low-energy behavior of this material is well captured by a half-filled extended Hubbard model, which supports the formation of Hubbard excitons (see Fig.~\ref{fig:fig1}a-b). These are comprised of doublon (doubly-occupied Cu site) and holon (empty Cu site) pairs, which can form symmetric (even) and antisymmetric (odd) superpositions, as confirmed by photoconductivity, electroreflectance, and infrared reflectivity measurements \cite{Kishida2000gigantic,ono2004linear,kim2008bound,kim2009optical}. Odd states are optically active (bright) with a lifetime of the order of 2~ps, while even excitons remain dark \cite{ogasawara2000ultrafast,ono2004linear}. Unlike in conventional semiconductors, the onsite Coulomb repulsion ($U$) causes the excitonic wavefunction to develop a node at the origin of relative holon-doublon coordinate, resulting in near-degenerate states separated by at most 16 meV \cite{kim2009optical}. Figure \ref{fig:fig1}c presents the frequency-dependent reduced optical conductivity $\omega\sigma_1(\omega)$ of Sr$_2$CuO$_3$ along the $b$ axis, as extracted with broadband infrared spectroscopy (see \textit{Methods}). All our optical measurements are conducted at room temperature, which is small compared to spin-exchange energy, but well above magnetic ordering temperature, ensuring the one-dimensional character of the electronic response. The excitonic states lie just below the $1.8$~eV charge-transfer gap, and exhibit minimal temperature dependence~\cite{kim2008bound}. At higher energies, the optical absorption reveals a continuum of unbound holon-doublon excitations~\cite{Essler2001excitons,kim2009optical}. We fit the experimental $\omega\sigma_1(\omega)$ using an extended Hubbard model in the large Mott gap limit (Supplementary  Section 1) to extract the Hubbard parameters~\cite{Essler2001excitons,Jeckelmann2003optical}.

Owing to quantum confinement, Hubbard excitons in one-dimensional Mott insulators exhibit large optical nonlinearities \cite{Kishida2000gigantic,mizuno2000nonlinear,kishida2001large,maeda2004third}, which are uniquely suited to optically probe and control their quantum states. In centrosymmetric Sr$_2$CuO$_3$, the lowest observable nonlinearity is of the third order and highly directional along the chain axis. The third-order susceptibility shows a colossal enhancement ($\chi^{(3)}(-3\omega; \omega, \omega, \omega)\approx1.4\cdot 10^{7}\:\mathrm{pm^2/V^2}$) at the $3\omega$ resonance with the Hubbard exciton, driven by the strong dipole coupling between odd- and even-parity states \cite{kishida2001large}. As a result, the chains generate third-harmonic photons at the exciton energy upon in-gap excitation. Since the THG yield critically depends on the transition dipoles $\vec{\mu}_{0u}=\langle 0| e\hat{\vec{r}}|u\rangle$ and $\vec{\mu}_{ug}=\langle u| e\hat{\vec{r}}|g\rangle$ coupling ground state ($|0\rangle$), odd ($|u\rangle$) and even ($|g\rangle$) exciton configurations, THG optical measurements probe the Hubbard exciton symmetry with a sensitivity beyond that of linear absorption (Fig.~\ref{fig:fig1}c-d).

\section*{Quantum control of the many-body wavefunction}

\begin{figure}[H]
    \centering
    \includegraphics[width = \textwidth]{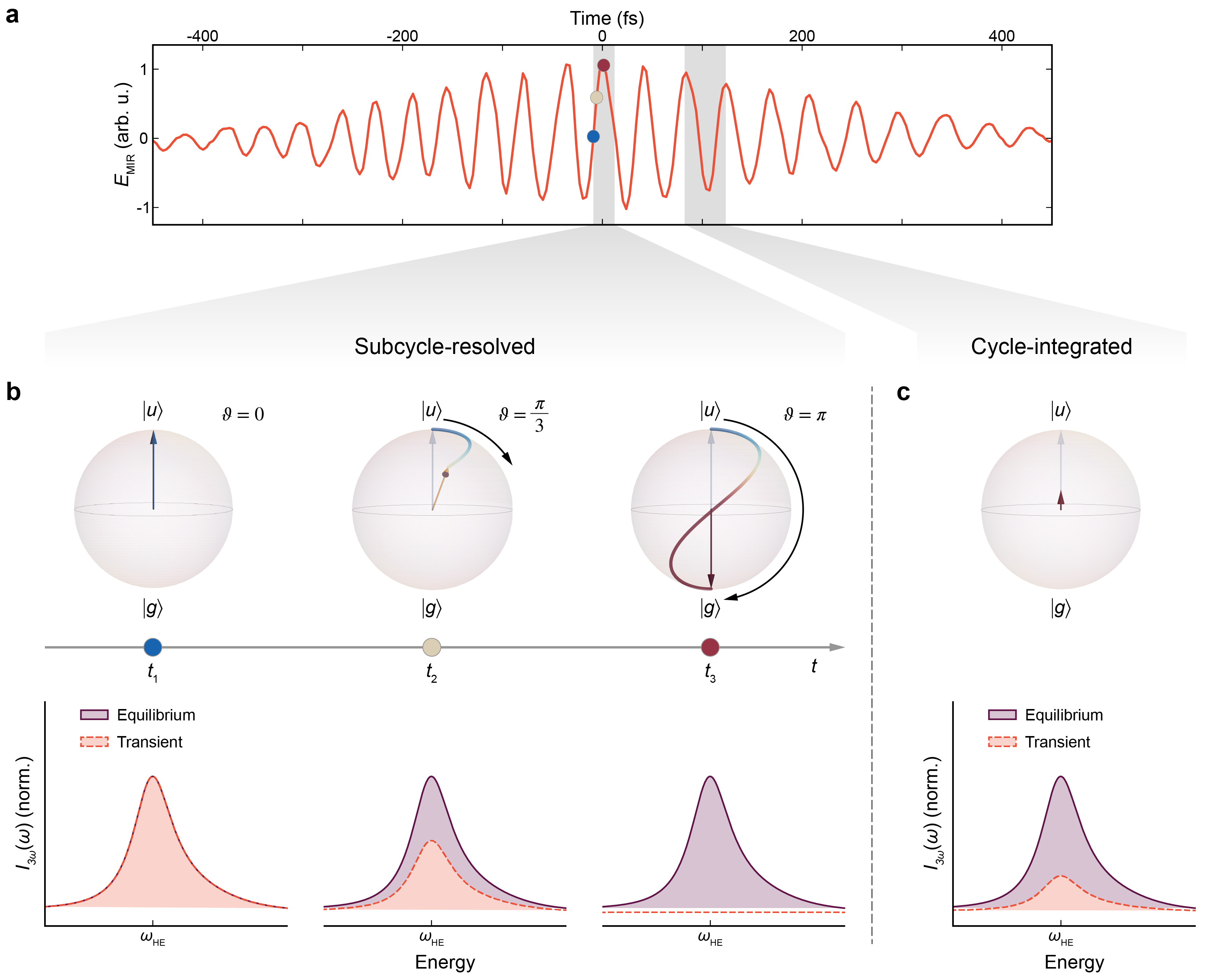}
    \caption{\label{fig:fig2} \textbf{Quantum control of the Hubbard exciton.} \textbf{a.} Experimental midinfrared pump field (center energy 0.12~eV, $\Delta\omega/\omega\sim8\%$). \textbf{b.} The instantaneous pump field at each time delay mixes even- ($|g\rangle$) and odd-parity ($|u\rangle$) exciton wavefunctions, initiating a subcycle rotation of the quantum state on the Bloch sphere. Sufficiently strong pump fields can coherently and instantaneously rotate the wavefunction by angles of $\vartheta\simeq\pi$. The energies of the two nearly-degenerate states remain unperturbed to first order, and the Bloch vector length is preserved during the rotation. As the quantum state rotates, the subcycle third-harmonic yield decreases due to changes in the dipole matrix elements, vanishing at $\vartheta=\pi$. \textbf{c.} The cycle-integrated rotation corresponds to a reduction of the Bloch vector length and a partial suppression of the third-harmonic emission.}
\end{figure}

In this work, we use intense midinfrared (MIR) pulses to coherently control the Hubbard exciton (Fig.~\ref{fig:fig2}a). Separated from the holon–doublon continuum, the two excitonic states constitute a many-body analogue of a quantum-optical two-level system. The relevant interaction Hamiltonian is $\hat{H}_{\mathrm{int}}(t)=\vec{\mu}_{ug}\cdot \vec{E}_\mathrm{MIR} ~|u\rangle\langle g| + \textrm{h.c.}$, where $\vec{E}_\mathrm{MIR}(t)=\vec{E}_\mathrm{MIR}^0\cos(\Omega t)$ is the pump field with frequency $\Omega$. Optical driving generally introduces both energy shifts and mixing of the quantum states. The shifts arise from optical Stark and Bloch–Siegert terms with opposite signs \cite{Sie2015valley,Sie2017large,shan2021giant}, often considered key signatures of Floquet driving. However, near degeneracy these terms largely cancel, leaving state mixing as the dominant optical perturbation. The Floquet wavefunctions $|u(t)\rangle$ and $|g(t)\rangle$ thus retain the same quasienergy, but coherently mix even and odd components \textit{within} the pump cycle:
\begin{align}
|u(t)\rangle &= e^{i E_{\rm exc} t/
\hbar} \left[ \cos\frac{\vartheta(t)}{2}|u\rangle+e^{i\varphi(t)}\sin\frac{\vartheta(t)}{2}|g\rangle \right] \\
|g(t)\rangle &= e^{i E_{\rm exc} t/
\hbar} \left[ \cos\frac{\vartheta(t)}{2}|g\rangle+e^{i\varphi(t)}\sin\frac{\vartheta(t)}{2}|u\rangle \right] .
\end{align}
This phenomenon corresponds to a coherent rotation of the wavefunction on a Bloch sphere spanned by even and odd exciton states, with $\vartheta(t)$ and $\varphi(t)$ as time-dependent spherical angles evolving at twice the pump frequency. Averaged over a pump cycle, the Floquet exciton state $\ket{u(t)}$ has reduced odd-parity character, redistributing spectral weight into Floquet sidebands at $E_{\rm exc} \pm 2n\Omega$, where $n$ is an integer. The even-parity state $\ket{g(t)}$ similarly admixes odd-parity components into sidebands at $E_{\rm exc} \pm (2n+1)\Omega$ (see \textit{Methods}).

We probe the evolving state via time-resolved resonant THG. The formation of Floquet-Hubbard excitons can be equivalently understood as Rabi oscillations between $|u\rangle$ and $|g\rangle$ states. In-gap NIR pulses create a population of odd-parity excitons ($|u\rangle$ state) through three-photon absorption. The MIR field drives Rabi oscillations between $|u\rangle$ and $|g\rangle$, transferring part of the population into the even-parity state. As $|g\rangle$ is optically dark for third-harmonic emission, this coherent population redistribution leads to a reduction of the THG response at the original $3\omega$ resonance when averaged over the probe pulse width. Because the MIR pump is off-resonant from the near-degenerate exciton states and the light–matter coupling $g = \vec{\mu}_{ug} \cdot \vec{E}_\mathrm{MIR}$ is comparable to $\Omega$, the Rabi frequency locks to the drive frequency, with $g$ setting only the rotation amplitude (Supplementary Section 4.2.1). This regime departs significantly from conventional resonant two-level driving and previous experiments in semiconductor quantum dots \cite{Cundiff1994rabi,Cole2001coherent,Press2008complete,Berezovskt2008picosecond}.

To visualize the connection between state parity and the THG signal, we sketch the time-dependent evolution of the Bloch vector and an instantaneous THG yield $I_{3\omega}(\omega)$ of the Hubbard exciton (Fig.~\ref{fig:fig2}b; see \textit{Methods}). In equilibrium, THG emission peaks at the exciton resonance and predominantly arises from a dipole transition between the ground state and the bright odd-parity exciton. As the MIR pump field oscillates, the many-body state (populated by the NIR probe) rotates on the Bloch sphere, returning to its initial orientation with each pump period. At zero field, the exciton population remains in the odd state; with increasing pump field, Rabi oscillations progressively cycle more population to the even state, reducing the THG signal at the original resonance. If the state is rotated by $\vartheta = \pi$, the third-harmonic emission at $3\omega = E_{\rm exc}$ vanishes. Upon cycle averaging, these Bloch vector dynamics lead to a reduced vector length and partial suppression of the THG yield (Fig.~\ref{fig:fig2}c).

\section*{Floquet control of the Hubbard exciton}

\begin{figure}[H]
    \centering
    \includegraphics[width = 0.8\textwidth]{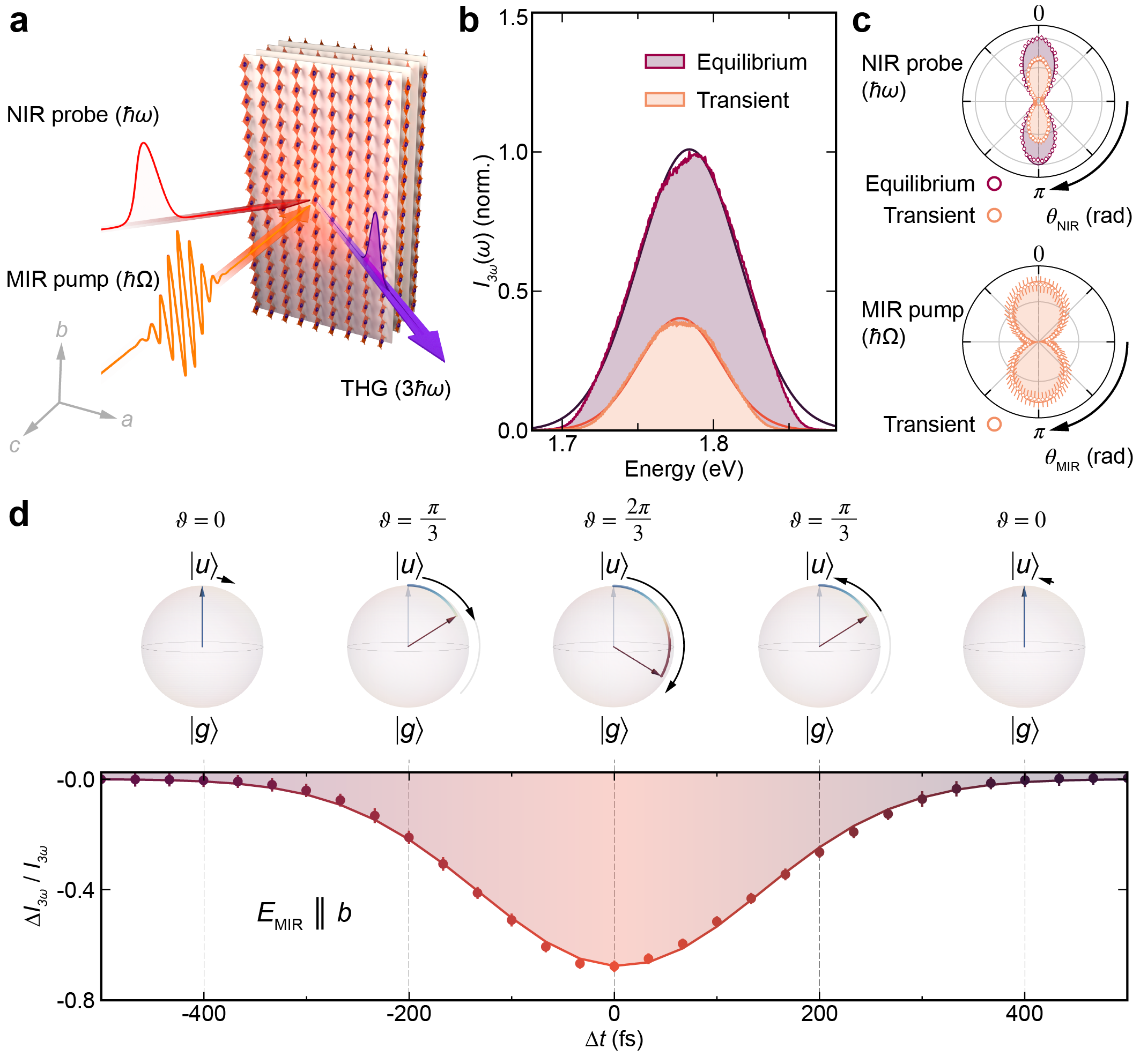}
    \caption{\label{fig:fig3} \textbf{Detecting the wavefunction rotation via third harmonic renormalization.} \textbf{a.} Schematics of the experimental setup. A MIR pump (orange, $\hbar\Omega$) excites the sample at normal incidence, while a NIR probe (red, $\hbar\omega$) impinges at $60^\circ$ to generate a third harmonic (THG, purple) measured in reflection. All beams are polarized parallel to the chains ($E \parallel b$). \textbf{b.} Third harmonic spectra of a $0.59$~eV NIR probe at equilibrium (purple) and after a $0.12$~eV pump ($E_\mathrm{MIR}=1.8$~MV/cm, orange, $\Delta t=0$~fs). Spectra are normalized to the equilibrium maximum and fit to Gaussian profiles (thin solid lines). \textbf{c.} THG patterns as a function of NIR probe (top) and MIR pump (bottom) polarization angles. The NIR probe is kept fixed at 0.59 eV, while $E_{\mathrm{MIR}}=1.2$~MV/cm. The MIR polarization data is symmetrized by averaging the two branches $[0,\pi]$ and $[\pi,2\pi]$. Error bars are standard deviations, while the solid line is a $1-J_0(A_1\cos\theta)^4$ fit with $A_1=(0.597 \pm 0.010)$. \textbf{d.} Time-resolved differential THG intensity (circles) for $E_\mathrm{MIR}=1.8$~MV/cm, fit to a $322.2(2.2)$~fs Gaussian profile. Error bars represent one standard deviation. Dashed lines mark time delays at which we calculate the instantaneous pump-induced angular rotation $\vartheta$ of Hubbard excitons with a three-level model.}
\end{figure}

We experimentally demonstrate quantum control of the Hubbard exciton via nonresonant midinfrared driving. We center the pump spectrum at $0.12$~eV ($28$~THz), well below the Mott gap and Hubbard exciton resonance, to prevent the generation of incoherent quasiparticles. At the same time, the pump energy is high enough to suppress quantum tunneling and the direct excitation of Sr$_2$CuO$_3$ phonons~\cite{Walters2009effect}. As shown in Fig.~\ref{fig:fig3}a, MIR pump pulses impinge on the sample at normal incidence, while the NIR probe is focused at a finite angle to enable the spectrally-resolved detection of THG emission in reflection geometry. All electric fields are linearly polarized along the chain direction ($b$ axis) of a Sr$_2$CuO$_3$ single crystal. At equilibrium, NIR pulses centered around 0.59 eV give rise to an intense THG emission at the three-photon resonance with the Hubbard exciton levels (see Fig.~\ref{fig:fig3}b). By rotating the NIR probe polarization, we verify that the THG emission pattern is $C_2$ symmetric and parallel to the chain, thus reflecting the one-dimensional character of the  on-chain electronic states (Fig.~\ref{fig:fig3}c). Upon excitation with $1.8$~MV/cm MIR fields (Fig.~\ref{fig:fig3}b), the THG emission undergoes a dramatic suppression (up to 70\%). Equilibrium and transient THG emission exhibit the same rotational pattern (Fig.~\ref{fig:fig3}c), thus indicating that the MIR excitation does not introduce multipole moments in the exciton symmetry. We also note that the THG energy changes at most by 7 meV, indicating a negligible pump-induced shift of both excitonic energy levels (Supplementary Section 6).

The THG intensity suppression is consistent with a coherent Floquet dressing and rotation of the Hubbard exciton wavefunction. By varying the MIR pump polarization, we find that the differential THG signal vanishes when the driving electric field is polarized normal to the chain (Fig.~\ref{fig:fig3}c). This polarization dependence fits well with a $1-J_0\big(\frac{\mu E_{\mathrm{MIR}}\cos\theta}{\hbar\omega}\big)^4$ dependence ($J_0$ is the zeroth Bessel function), expected from a Floquet-dressing of the excitonic states (Supplementary Section 4). The time-dependent THG intensity suppression (Fig.~\ref{fig:fig3}d) is symmetric about zero delay, exhibiting a Gaussian profile with a 320-fs width. This behavior follows the intensity convolution of the 270-fs MIR pump and 80-fs NIR probe, as expected in a Floquet-driven response. We note that in our experiment the NIR probe duration effectively averages the Bloch vector rotation over 2–3 cycles. Absorption of three probe photons prepares the exciton in the $|u\rangle$ (odd) state. As the mid-infrared field grows, the exciton wavefunction then rotates from $|u\rangle$ to $|g\rangle$ with an increasing angle, suppressing the third-order nonlinearity. Upon removal of the driving field, the THG yield returns to its unperturbed value. For negligible spin-orbit coupling of the valence holes on Cu and linear polarization along the chain, $\vec{\mu}_{ug} \cdot \vec{E}_\mathrm{MIR}$ is real and $\varphi(t) = 0$; the remaining angle can be found exactly $\vartheta(t) = 2\arccos\left[ \sum_m J_{2m}\left(\frac{\vec{\mu}_{ug} \cdot \vec{E}_{\mathrm{MIR}}}{\hbar\Omega}\right)e^{2im\Omega t}\right]$ where $J_m$ are Bessel functions (see Methods). For a pump field of 1.8 MV/cm and a dipole matrix element $\mu_{ug}=7.37$~\AA~\cite{Kishida2000gigantic}, we theoretically anticipate Floquet-dressed exciton states with a peak instantaneous rotation angle of $2\pi/3$ (Fig.~\ref{fig:fig3}d). 

\begin{figure}[H]
    \centering
    \includegraphics[width = \textwidth]{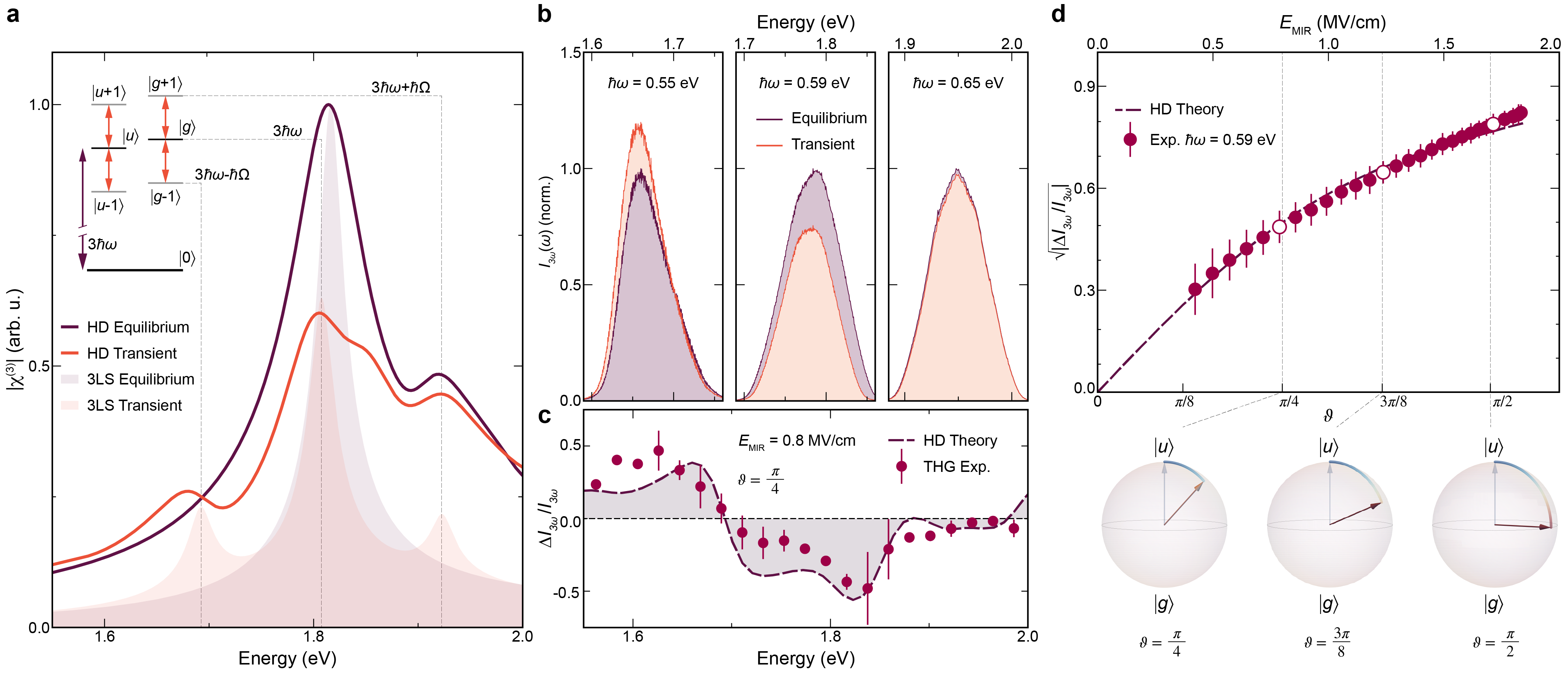}
    \caption{\label{fig:fig4} \textbf{Floquet sidebands of the Hubbard exciton.}
    \textbf{a.} Periodic MIR driving of even- and odd-parity states generates Floquet replicas spaced by $\hbar\Omega$, which produce sidebands in the nonlinear response. Third-order susceptibility in- (purple) and out-of-equilibrium (orange) for a three-level system (3LS) including the ground state and two excitonic levels, alongside a full holon-doublon (HD) model incorporating holon-doublon continuum and experimental broadening. Driven susceptibility is computed for a MIR pump field $E_\mathrm{MIR}=1.85$~MV/cm.
    \textbf{b.} Third harmonic spectra for selected NIR probe energies at equilibrium (purple) and after a $0.12$~eV pump ($E_\mathrm{MIR}=0.8$~MV/cm, orange, $\Delta t=0$~fs). \textbf{c.} Experimental (red circles) and theoretical (HD, dashed purple line) differential THG intensity at fixed MIR field $E_\mathrm{MIR}=0.8$~MV/cm, showing suppression of the main THG resonance and emergence of a below-gap Floquet sideband. Experimental spectra are binned to 20 meV, and error bars are standard deviations of 7 independent measurements. \textbf{d.} Differential third-order nonlinearity ($\sqrt{|\Delta I_{3\omega} / I_{3\omega}|}\propto\Delta\chi^{(3)} / \chi^{(3)}$) at $\hbar\omega=0.59$~eV as a function of MIR field and instantaneous rotation angle from the HD Floquet model (red circles). White circles mark the special angles $\vartheta=\pi/4,\: 3\pi/8,\: \mathrm{and}\: \pi/2$.}

    \end{figure}

To further validate the Floquet mechanism, we measure the energy dependence of the third-order nonlinearity.
Periodic driving generates sidebands from transitions involving Floquet replicas of excitonic levels (Fig.~\ref{fig:fig4}a). We theoretically evaluate the third-order susceptibility $\chi^{(3)}$ of Sr$_2$CuO$_3$ using both a simplified three-level model and a Floquet holon-doublon Hamiltonian capturing the full continuum above the Mott gap (see Methods, and Supplementary Section 4). The applicability of these electronic models is validated by complementary time-dependent exact-diagonalization calculations of the dynamical third harmonic response within the extended Hubbard model (Supplementary Section 3). In the three-level approximation, the equilibrium third-order response is sharply peaked at the exciton energy. Under periodic driving, the main peak is suppressed, and first-order sidebands emerge at $\pm\hbar\Omega$. Accounting for the continuum and realistic broadening reduces sideband visibility, particularly above the Mott gap, where dressed states scatter with the holon-doublon continuum. However, the sub-gap sideband persists, leading to off-resonant enhancement of the third harmonic signal.

We probe this non-monotonic behavior by detuning the NIR probe. We scan the THG response across the Hubbard exciton resonance by varying the NIR energy from $0.55$ to $0.66$~eV, while keeping MIR pump energy and field ($0.8$~MV/cm) fixed. For each probe energy, we record equilibrium and transient THG spectra and extract the differential third harmonic intensity (see Fig.~\ref{fig:fig4}b and Supplementary Section 2). Above the Mott gap, the THG intensity remains nearly unchanged. At resonance with the Hubbard exciton, the signal is suppressed by 50\%. Below the gap, the THG is enhanced by 40\%, consistent with the emergence of a Floquet sideband in $\chi^{(3)}$. We can rule out several alternative explanations for the intensity enhancement. First, the sideband cannot arise from a fourth-order nonlinearity, which is forbidden in this centrosymmetric compound. Second, it cannot result from the small pump-induced energy shifts of the Hubbard exciton states ($\sim 4$ meV at 0.8 MV/cm) \cite{Sie2017large}, which would produce a differential THG intensity with a zero near 1.8 eV. Third, although sum-frequency peaks appear away from the resonance, their MIR field dependence excludes an intensity tradeoff between competing nonlinear processes as the source of the modulation (Supplementary Section 5). Rather, as shown in Fig.~\ref{fig:fig4}c, the observed sideband quantitatively agrees with the theoretical harmonic yield of a Floquet holon-doublon model with a maximum Bloch sphere rotation angle of $\vartheta=\pi/4$, confirming the coherent dressing of the exciton states as the most natural explanation of our data.

We finally demonstrate rotations of the Hubbard exciton wavefunction by arbitrary angles. By tuning the incident MIR field, we measure the field scaling of the THG emission at $\hbar\omega=0.59$~eV.
Figure~\ref{fig:fig4}d shows the field dependence of the differential third-order nonlinearity, $\Delta\chi^{(3)} / \chi^{(3)} \propto\sqrt{|\Delta I_{3\omega} / I_{3\omega}|}$. Comparing the experimental nonlinearity to the Floquet holon-doublon model yields an accurate estimate of the subcycle rotation angle. The Hubbard exciton continuously rotates in an oscillatory fashion with increasing field, reaching a peak instantaneous angle $\vartheta=\pi/2$ at $1.7$~MV/cm. These values slightly undershoot predictions from a three-level model, reflecting the influence of field interactions with unbound holon-doublon states. Nevertheless, the exceptionally large rotation angles achieved under off-resonant excitation conditions highlight the colossal susceptibility of Hubbard excitons to external electric fields.
Figure~\ref{fig:fig4}d shows the field dependence of the differential third-order nonlinearity, $\Delta\chi^{(3)} / \chi^{(3)} \propto\sqrt{|\Delta I_{3\omega} / I_{3\omega}|}$. Comparing the experimental nonlinearity to the Floquet holon-doublon model yields an accurate estimate of the subcycle rotation angle. The Hubbard exciton continuously rotates in an oscillatory fashion with increasing field, reaching a peak instantaneous angle $\vartheta=\pi/2$ at $1.7$~MV/cm. These values slightly undershoot predictions from a three-level model, reflecting the influence of field interactions with unbound holon-doublon states. Nevertheless, the exceptionally large rotation angles achieved under off-resonant excitation conditions highlight the colossal susceptibility of Hubbard excitons to external electric fields.

\section*{Conclusion and outlook} 

In conclusion, we demonstrated quantum control of Hubbard excitons in a prototypical Mott insulator. By applying ultrafast off-resonant fields and probing the many-body state via resonant third-harmonic generation, we observed a coherent Floquet renormalization of the third-order Hubbard exciton response, consistent with $U(1)$ control of the many-body wavefunction. Upon varying the pump field, the exciton state undergoes controllable rotations by arbitrary angles, reaching and exceeding $\pi/2$ as a function of electric field strength and revealing a high degree of tunability. This establishes a novel Floquet engineering regime in which wavefunction mixing dominates, realizing a Rabi problem at ultrastrong coupling (Supplementary Section 4.2.1).

By demonstrating the coherent manipulation of an addressable many-body excitation in a paradigmatic strongly-correlated electron system, our results establish a platform for ultrafast quantum control in solids. The combination of optimal control protocols \cite{Rabitz2004quantum} and midinfrared pulse shaping \cite{Kaindl2000generation,Cartella2014pulse} offers a route to tailoring more complex Bloch-sphere trajectories and driving long-lived population transfer between quantum states. Extension to the terahertz regime will further reduce pump field requirements and enable resonant coupling between near-degenerate quantum states. Uniaxial $\pi/2$ rotations represent the minimal building block for programmable pulse sequences implementing Ramsey interferometry, dynamical decoupling, and Waugh-Huber-Haeberlen control schemes \cite{Vandersypen2005NMR,Choi2020robust}. The application of these methods to correlated few-state excitations with strong spin-orbit coupling will also enable wavefunction rotations in the azimuthal plane \cite{Kim2014excitonic}, thus achieving full $SU(2)$ control. These capabilities will be key for designing novel photoinduced states of matter \cite{Kunes2015excitonic,Bloch2022strongly} and achieving on-demand control of optical nonlinearities in quantum materials.

Finally, quantum control based on Floquet driving could find applications in quantum sensing applications. The exciton energy and lineshape are highly sensitive to electronic interactions and chemical potentials \cite{Xia2024optical}, while field-dependent wavefunction rotation amplitude enables probing of local dielectric screening. Controlled wavefunction rotations, especially in systems with narrower excitonic linewidths, such as Rydberg excitons in Cu$_2$O \cite{Kazimierczuk2014giant} and magnetically-coupled excitons in van der Waals materials \cite{Kang2020coherent,Bae2022exciton}, could be used to implement decoupling and recoupling sequences that enhance sensitivity and functionalize these systems as quantum sensors in a variety of quantum materials.

\clearpage

\clearpage
\singlespacing
\bibliography{floquet_sr2cuo3}

\doublespacing
\section*{Methods}
\subsection*{Sample preparation and characterization}

High-quality single crystals of Sr$_2$CuO$_3$ were grown using the traveling solvent-floating zone method. Powders of SrCO$_3$ and CuO (99.99\%) in their metal ratio were mixed, ground, and calcined (24 hrs at 980~$^{\circ}$C, 48~hrs at 1050~$^{\circ}$C for feed rods, and 950~$^{\circ}$C for Sr$_2$CuO$_3$ solvent materials). Powders were reground, placed into a rubber tube, and hydrostatically pressed at 4200~Kg/cm$^2$ to form feed rods. The pressed feed rods (8~mm diameter, 25~cm length) were sintered for 72~hrs at 1100~$^{\circ}$C in air. Sintered rods of the oxide powder Sr$_2$CuO$_3$ were prepared as feed rods for single crystal growth. We used a floating zone furnace equipped with two ellipsoidal mirrors to grow the single crystals. The sintered feed and seed rods were contra-rotated to achieve homogeneous heating of the floating zone and promote mixing in the zone. The growth velocity was 1~mm/hour with 1~bar oxygen pressure to obtain large single crystal rods ($\sim8$~mm diameter, $\sim10$~cm length). The rod was then cut into smaller crystals with a typical size of $2\times2\times2$~mm$^3$. We characterized the samples using x-ray diffraction and x-ray absorption spectroscopy (Cu $L_3$- and O $K$-edges) and verified the agreement with previously published results. Before optical and x-ray absorption spectroscopy measurements, the samples were freshly cleaved to expose the $ab$ plane and immediately transferred into vacuum to a pressure better than $5\cdot10^{-6}$~mbar to prevent surface degradation. The results presented here were reproduced in three different samples.

\subsection*{Equilibrium infrared optical measurements}
The equilibrium optical properties of Sr$_2$CuO$_3$ single crystals were measured using Fourier-transform infrared spectroscopy. Reflectivity at near-normal incidence was recorded along the chain direction of freshly cleaved samples, utilizing a Bruker 66 v/S Fourier transform spectrometer for energies up to 3~eV and a Cary 5 grating spectrometer for the 3–6 eV range \cite{kim2008bound}. To ensure accurate measurements, optical spectra were referenced in-situ via gold evaporation, and a double polarization configuration was employed to suppress spurious polarization effects from reflections on the sample and mirrors. The optical conductivity spectrum was determined through Kramers-Kronig transformation of the measured reflectivity $R(\omega)$.

\subsection*{Time-resolved THG measurements}\label{sec:setup}
We used a Ti:sapphire amplifier operating at 1 kHz (7~mJ pulse energy, 35~fs pulse duration, 800~nm wavelength) to generate MIR pump and NIR probe light. The amplifier output was directed into a TwinTopas optical parametric amplifier to generate NIR beams centered at $1.29\ \mathrm{\mu m}$ and $1.43\ \mathrm{\mu m}$. Carrier-envelope-phase-stable MIR pulses were obtained by difference frequency generation (DFG) between the two NIR beams in a $600~\mathrm{\mu m}$-thick GaSe crystal. The MIR light was fixed at $28$~THz with a $\sim300$-fs pulse duration. The MIR pulses were characterized via electro-optical sampling in a 30~$\mu$m GaSe crystal using sub-10-fs visible (VIS $\approx 520$~nm) pulses from a non-collinear optical parametric amplifier (NOPA) as a gate. After generation, the MIR light was recollimated and focused onto the sample at normal incidence. The beam waist had a $300\ \mathrm{\mu m}$ diameter, leading to electric fields up to $\sim 1.8$~MV/cm. The MIR field strength was tuned by propagating the beam through an Altechna MgF$_2$ waveplate and wire-grid polarizer. A portion of the NIR idler was used as a THG driver. The NIR beam propagated at an angle of $30^\circ$ with respect to the surface normal and was focused down to a 75~$\mathrm{\mu m}$-diameter spot to achieve peak fields below 0.9 MV/cm. The MIR penetration depth is $100$~nm, while that of the NIR pulse is 11~nm, meaning that the probe samples a homogeneously excited volume. The third-harmonic light emitted by the sample is measured in reflection geometry. We recorded time- and energy- resolved THG data in the range $1.24-5.90$~eV ($210-1000$~nm) at each pump-probe delay using an Ocean Optics triggered spectrometer. At each pump-probe delay, we recorded spectra with and without MIR pump in an alternating pattern. At a given pump-probe delay $\Delta t$, we extracted the differential THG yield $\Delta I_{3\omega} / I_{3\omega}$  by integrating the experimental yield:
\begin{equation*}
    \frac{\Delta I_{3\omega}}{I_{3\omega}} \equiv \frac{ \int_{\omega_\mathrm{min}}^{\omega_\mathrm{max}} I_\mathrm{tr}(\omega; \Delta t)-I_\mathrm{eq}(\omega)\mathrm{d}\omega}{\int_{\omega_\mathrm{min}}^{\omega_\mathrm{max}} I_\mathrm{eq}(\omega)\mathrm{d}\omega},
\end{equation*}
where $I_\mathrm{eq}$ and $I_\mathrm{tr}$ are the equilibrium and transient (THG) spectra. The integration boundaries $\omega_\mathrm{min}$ and $\omega_\mathrm{max}$ are set to encompass the spectral regions where the THG intensity is within 90\% of the equilibrium THG maximum $I_\mathrm{eq}^\mathrm{max}$, i.e., $I_\mathrm{eq}(\omega_\mathrm{min/max})/I_\mathrm{eq}^\mathrm{max}=0.1$.

\subsection*{Floquet holon-doublon theory}

The large optical nonlinearity of Mott insulating chains arises from states which can be described by an effective holon-doublon (HD) model \cite{mizuno2000nonlinear,kishida2001large} driven by light $\hat{H}(t) = \hat{H}_{\rm hd} + \hat{P}_{\rm dip} \mathcal{E}(t)$ where
\begin{align}\label{eq:HD}
    \hat{H}_{\rm hd} &= -t \sum_i \left ( \hat{h}^{\dagger}_{i+1} \hat{h}_i + \hat{d}^{\dagger}_{i+1} \hat{d}_i + h.c. \right)
    - V \sum_{\left < i,j
    \right > } \hat{h}^{\dagger}_i \hat{h}_i \hat{d}^{\dagger}_j \hat{d}_j + \frac{U}{2} \sum_i \left( 
    \hat{h}^{\dagger}_i \hat{h}_i 
    + 
    \hat{d}^{\dagger}_i \hat{d}_i
    \right) ~, \\
    \hat{P}_{\rm dip} &= e a_0 \sum_i i \left(  \hat{h}^{\dagger}_i \hat{h}_i - \hat{d}^{\dagger}_i \hat{d}_i \right). 
\end{align}
Here, $\hat{h}_i^{\dagger}$ ($\hat{d}_i^{\dagger}$) describe the creation of a holon (doublon) on site $i$ with energy $U$, and a constraint is imposed such that each site can host at most one doublon or holon. The hopping of holons and doublons is parametrized by $t$ (taken to be the same sign after a gauge transformation), and $V$ denotes an attractive interaction between holons and doublons on neighboring sites. The spin degrees of freedom are effectively decoupled due to spin-charge separation. Finally, $\hat{P}_{\rm dip} \mathcal{E}(t)$ describes coupling to light in the dipole gauge, where $e$ is the electron charge, $a_0$ is the Cu-Cu distance, and $\mathcal{E}(t) = \mathcal{E} \cos(\Omega t)$ is the pump (MIR) electric field with frequency $\Omega$. Since the pump is far below resonance with the charge gap, the equilibrium ground state is not appreciably changed by the drive and photoexcitation of holon-doublon pairs by pump photons can be neglected. In contrast, if holon-doublon pairs are generated (via absorbing probe photons), then the resulting excited states must become strongly dressed by the pump.

The third-harmonic pump-probe response can now be computed straightforwardly via the Floquet formalism. As we are focused on the features near the exciton peak, we constrain the Hamiltonian to the sector with a single holon-doublon pair. We now compute the Floquet eigenstates $\ket{\Psi_n(t)} = e^{-i\varepsilon_nt/\hbar } \sum_m e^{im\Omega t} \ket{\Phi_{n,m}}$ of $\hat{H}(t)$ via finding the eigenstates $\ket{\Phi_{n,m}}$ and quasi-energies $\varepsilon_n$ of the Sambe space Hamiltonian $\hat{H}_F = \sum_m \left( \hat{H}_{\rm hd} + m \hbar\Omega \right) \otimes \ket{m}\bra{m} + \sum_m \hat{P}_{\rm dip} \otimes \left( \ket{m+1}\bra{m} + \ket{m}\bra{m+1} \right)$, using $40$ Floquet sidebands for convergence. The third-harmonic response can now be computed via a spectral representation in Floquet basis (Supplementary Section 4). Up to scaling factors, we find
\begin{align} \label{eq: THG_floquet}
    \chi^{(3)}(-3\omega; \omega,\omega,\omega) &\propto \sum_{\gamma\mu\nu}\sum_{m_1m_2m_3m_4} \sum_{l_1+l_2+l_3+l_4=0} \muvec_{0 \gamma}^{m_1,m_1+l_1} \muvec_{\gamma\mu}^{m_2,m_2+l_2} \muvec_{\mu\nu}^{m_3,m_3+l_3}\muvec_{\nu 0}^{m_4,m_4+l_4} \times \notag\\
     & \times \left(  \frac{1}{( \Omega_{\gamma 0} - \Omega^{l_2l_3l_4}  - 3\omega)(  \Omega_{\mu 0} - \Omega^{l_3l_4} - 2\omega)(  \Omega_{\nu 0} - \Omega^{l_4} -  \omega)}  + \right. \notag\\
    & + \frac{1}{( \Omega_{\gamma 0}^\ast - \Omega^{l_2l_3l_4}  + \omega )(  \Omega_{\mu 0} - \Omega^{l_3l_4} - 2\omega)(  \Omega_{\nu 0} - \Omega^{l_4} - \omega)}  +  \notag\\
     & + \frac{1}{( \Omega_{\gamma 0}^\ast - \Omega^{l_2l_3l_4}  + \omega )(  \Omega_{\mu 0}^\ast - \Omega^{l_3l_4} + 2\omega)(  \Omega_{\nu 0} - \Omega^{l_4} -  \omega)}  + \notag\\
     & + \left. \frac{1}{( \Omega_{\gamma 0}^\ast - \Omega^{l_2l_3l_4}  + \omega )(  \Omega_{\mu 0}^\ast - \Omega^{l_3l_4} + 2\omega)(  \Omega_{\nu 0}^\ast - \Omega^{l_4} + 3\omega)}\right), 
\end{align}
where we define $\Omega^{l_1l_2\ldots l_k} = (l_1+l_2+\ldots + l_k)\Omega$ and $\Omega_{\alpha\beta} = \varepsilon_\alpha/\hbar - \varepsilon_\beta/\hbar - i\gamma$. Here, $\gamma$ is a phenomenological broadening factor. The dressed dipole moment connecting individual Floquet sidebands is defined as $\muvec_{\alpha\gamma}(t) = \bra{\Phi_\alpha(t)}\muvec\ket{\Phi_\gamma(t)} = \sum_{mn}e^{i(m-n)\Omega t}\bra{\phi_\alpha^m}\muvec \ket{\phi_\gamma^n} = \sum_{mn}e^{i(m-n)\Omega t}\muvec_{\alpha\gamma}^{mn}$.

The effective model parameters are chosen to align with the experiment. To compare simulated and experimental $\chi^{(3)}$, we tracked the suppression of the main peak intensity (integrated within a $\pm 0.05$~eV window around the main peak), yielding a parameter choice $t = 0.56$~eV, $V/t = 2.4$, and a broadening factor of $\hbar \gamma/t = 0.06$. Simulations were performed using a chain length of 16 sites.

\subsection*{Bloch sphere rotations and effective Floquet representation}

The quantum control of the third-harmonic response has an elegant effective Floquet representation if excitations into the holon-doublon continuum are integrated out. In a holon doublon theory, only the excitonic bound states and the ground state remain. Since the pump is far off-resonance with respect to the charge gap, the Floquet dressing of the ground state $\ket{0}$ can be neglected, and the effective Hamiltonian simplifies to a Rabi Hamiltonian for the two excitonic states. As the pump frequency is much larger than the anticipated finite splitting between excitons, we take $\ket{u}$ and $\ket{g}$ to be exactly degenerate with energy $\epsilon_{ex}$ with respect to the ground state, and obtain:
\begin{equation}
\hat{H}(t) = \epsilon_{ex} \left( \ket{u} \bra{u} + \ket{g} \bra{g} \right) + \mu_{ug} \mathcal{E} \cos(\Omega t)~ \left( \ket{u} \bra{g} + \ket{g} \bra{u} \right)~ .
\end{equation}
Here, $\mathcal{E}$ is the pump field polarized along the chain direction and $\Omega$ is its MIR frequency.

The Floquet eigenstates of $\hat{H}(t)$ can be found exactly (Supplementary Section 4.2). Importantly, while the \textit{static} inversion symmetry is broken on a subcycle scale, the Hamiltonian is characterized by a \textit{dynamical} inversion symmetry, given by the product of the static inversion operator and a time translation by half the MIR period. The dynamically even and odd Floquet eigenstates read:
\begin{align}
    \ket{\tilde{u}(t)} &= \sum_m e^{2im\Omega t} J_{2m}\left(\frac{\mu_{ug}\mathcal{E}}{\hbar\Omega}\right)\ket{u} + e^{i(2m+1)\Omega t} J_{2m+1}\left(\frac{\mu_{ug}\mathcal{E}}{\hbar\Omega}\right)\ket{g} ~, \\
    \ket{\tilde{g}(t)} &= \sum_m e^{2im\Omega t} J_{2m}\left(\frac{\mu_{ug}\mathcal{E}}{\hbar\Omega}\right)\ket{g} + e^{i(2m+1)\Omega t} J_{2m+1}\left(\frac{\mu_{ug}\mathcal{E}}{\hbar\Omega}\right)\ket{u} ~,
\end{align}
where $J_m$ denote Bessel functions. Substituting these into a Floquet-Lehmann representation of the third-harmonic response (Supplementary Section 4.1), one finds:
\begin{align}  \label{eq:THG_3S}
\chi^{(3)}(-3 \omega; \omega, \omega, \omega) ~\propto~
\sum_m
\frac{\left[J_m \left(\frac{\mu_{ug} \mathcal{E}}{\hbar\Omega} \right)\right]^2 \mu_{ug}^2 \mu_{0u}^2 }
{ (\epsilon_{ex}/\hbar + m \Omega - 3 \omega - i \gamma)
(\epsilon_{ex}/\hbar + m \Omega - 2 \omega - i \gamma)
(\epsilon_{ex}/\hbar + m \Omega - \omega - i \gamma) } ,
\end{align}
where $\mu_{0u}$ is the dipole transition matrix element between the ground state and odd excitonic state.

The coefficients of this effective theory can in principle be computed perturbatively from the holon-doublon model (Supplementary Section 4.3.1). We adopt a simpler procedure here: First, the exciton energy $\epsilon_{ex}$ is fixed by identifying the frequency of the main THG peak. Second, the effective dipole matrix element $\mu_{ug}$ is obtained by comparing the relative heights of the main peak and the side peak in the THG spectrum of the holon-doublon model and comparing to the effective model [Eq. (\ref{eq:THG_3S})].  Lastly, since $\mu_{u0}$ is an overall prefactor in Eq.~\eqref{eq:THG_3S}, its value is chosen to ensure that the height of the main peak matches that of the holon-doublon THG.

\section*{Acknowledgments}
We thank K. Burch, P. Cappellaro, A. Cavalleri, E. Demler, M. Eckstein, T. Giamarchi, D. Hsieh, H. Okamoto, D. Reis, T. Tohyama, P. Werner, and A. Yacoby for insightful discussions. This work was primarily supported by the U.S.\ Department of Energy, Office of Basic Energy Sciences, Early Career Award Program, under Award No.\ DE-SC0022883 (D.R.B. and M.M.) and Award No.\ DE-SC0024494 (D.C. and M.C.). D.C. acknowledges funding from the NSF GRFP under Grant No. DGE-1845298. The work performed at Brookhaven National Laboratory was supported by the U.S. Department of Energy, Division of Materials Science, under Contract No.~DE-SC0012704. We acknowledge funding from the Deutsche Forschungsgemeinschaft (DFG, German Research Foundation) - 531215165 (Research Unit ``OPTIMAL''). This work was supported by the Cluster of Excellence ‘Advanced Imaging of Matter' (AIM) and the Max Planck-New York City Center for Non-Equilibrium Quantum Phenomena. The Flatiron Institute is a division of the Simons Foundation. Simulations were performed with computing resources granted by RWTH Aachen University under projects rwth0752 and rwth1258. The authors gratefully acknowledge computing time on the supercomputer JURECA \cite{JURECA} at Forschungszentrum Jülich under grant no. enhancerg.

\section*{Author Contributions} D.R.B. and M.M.\ conceived the experiment. M.M.\ supervised the project. D.R.B. conducted the ultrafast optical measurements with support from F.G. and T.M. D. R. B. analyzed the data with help from all coauthors. C. C. H. characterized the equilibrium optical response of Sr$_2$CuO$_3$. D. C. and M. C. calculated the holon-doublon and three-level Floquet response. C. W.\ and I.-T. L. performed theoretical analyses and interpreted the data under the supervision of D. M. K., A. R., and M. M. Single crystals were synthesized by I. A. Z. and G. D. G. M.M. and M.C. wrote the manuscript with input from all other authors. 

\section*{Competing Interests}
The authors declare no competing interests.

\end{document}


\setlength{\parindent}{2ex}

\vspace{-4cm}

\title[Article Title]{\hspace{2.5cm}Supplementary Information for:\\ Quantum control of Hubbard excitons}

\author*[1,2]{\fnm{Denitsa R.} \sur{Baykusheva}}\email{denitsa.baykusheva@ista.ac.at}
\equalcont{These authors contributed equally to this work.}
\author[3]{\fnm{Deven} \sur{Carmichael}}
\equalcont{These authors contributed equally to this work.}
\author[4,5]{\fnm{Clara} \sur{S. Weber}}
\author[6]{\fnm{I-Te} \sur{Lu}}
\author[1]{\fnm{Filippo} \sur{Glerean}}
\author[1]{\fnm{Tepie} \sur{Meng}}
\author[1]{\fnm{Pedro B. M.} \sur{De Oliveira}}
\author[7]{\fnm{Christopher C.} \sur{Homes}}
\author[7]{\fnm{Igor A.} \sur{Zaliznyak}}
\author[7]{\fnm{G. D.} \sur{Gu}}
\author[7]{\fnm{Mark P. M.} \sur{Dean}}
\author[6,8]{\fnm{Angel} \sur{Rubio}}
\author[4,5,6]{\fnm{Dante} \sur{M. Kennes}}
\author*[3]{\fnm{Martin} \sur{Claassen}}\email{claassen@sas.upenn.edu}
\author*[1]{\fnm{Matteo} \sur{Mitrano}}\email{mmitrano@fas.harvard.edu}

\affil[1]{\orgdiv{Department of Physics}, \orgname{Harvard University}, \orgaddress{\city{Cambridge}, \state{MA}, \country{USA}}}

\affil[2]{\orgname{Institute of Science and Technology Austria}, \orgaddress{\city{Klosterneuburg}, \country{Austria}}}

\affil[3]{\orgdiv{Department of Physics and Astronomy}, \orgname{University of Pennsylvania}, \orgaddress{\city{Philadelphia}, \state{PA}, \country{USA}}}

\affil[4]{\orgdiv{Institut f\"ur Theorie der Statistischen Physik}, \orgname{RWTH Aachen University}, \orgaddress{\city{Aachen}, \country{Germany}}}

\affil[5]{\orgname{JARA-Fundamentals of Future Information Technology}, \orgaddress{\city{Aachen}, \country{Germany}}}

\affil[6]{\orgname{Max Planck Institute for the Structure and Dynamics of Matter, Center for Free-Electron Laser Science (CFEL)}, \orgaddress{\city{Hamburg}, \country{Germany}}}

\affil[7]{\orgdiv{Department of Condensed Matter Physics and Materials Science}, \orgname{Brookhaven National Laboratory}, \orgaddress{\city{Upton}, \state{NY}, \country{USA}}}

\affil[8]{\orgdiv{Initiative for Computational Catalysis (ICC) and Center for Computational Quantum Physics (CCQ)}, \orgname{Simons Foundation Flatiron Institute}, \orgaddress{\city{New York}, \state{NY}, \country{USA}}}

\date{\today}

\maketitle

\vspace{-10mm}
\tableofcontents
\doublespacing

\section{Quantifying the equilibrium Hubbard parameters}\label{sec:fit}
\begin{figure}[h]
\centering
\includegraphics[width=0.85\textwidth]{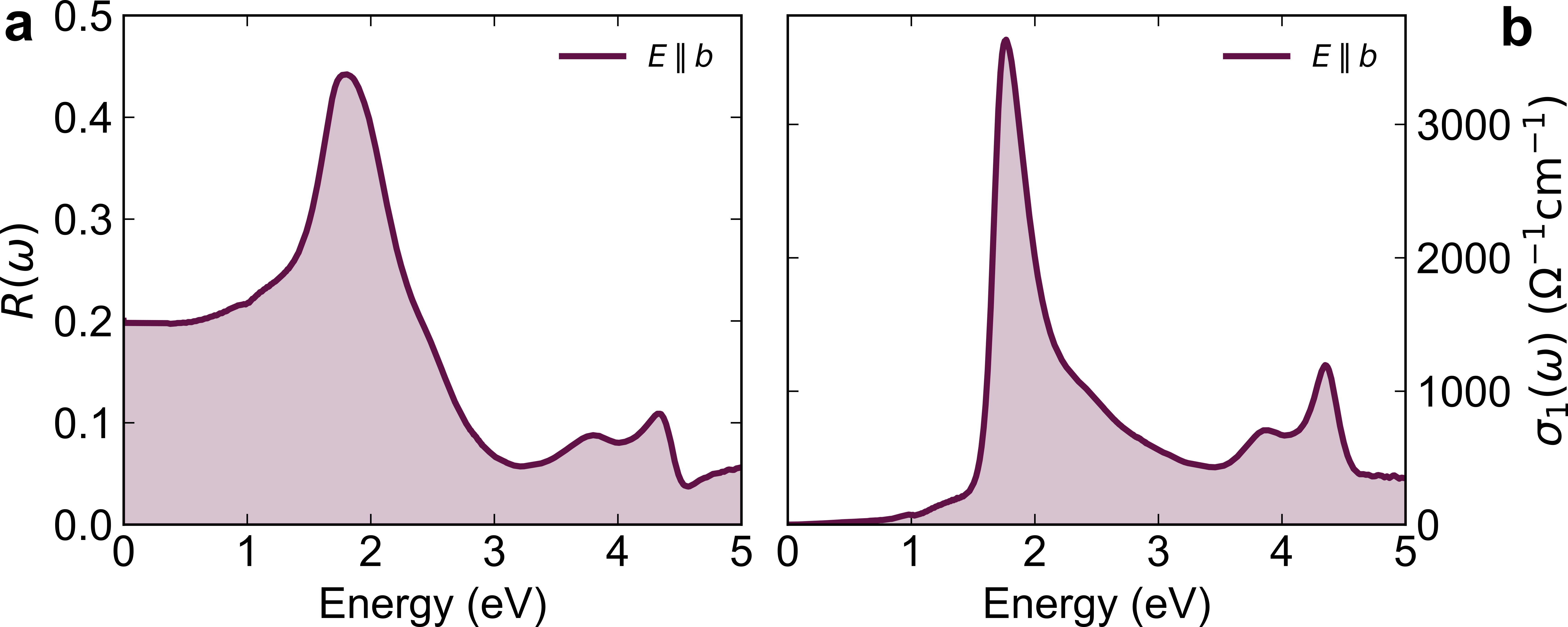}
\caption{\textbf{Equilibrium optical properties of Sr$_2$CuO$_3$.} \textbf{a.} Broadband reflectivity and \textbf{b.} optical conductivity of Sr$_2$CuO$_3$ at normal incidence and along the chain direction.}
\label{fig:KK}
\end{figure}

\vspace{-3mm}
Starting from the reflectivity $R(\omega)$, we extract the optical conductivity from 0 to $5.8$~ eV by standard Kramers-Kronig (KK) analysis \cite{wooten1972optical}. Figs.~\ref{fig:KK}a-b show the normal-incidence reflectivity and optical conductivity for light polarized along the chains and at 295 K. To determine the extended Hubbard model (EHM) parameters ($t$, $U$, $V$) best capturing the optical response of $\mathrm{Sr_2CuO_3}$, we fit the experimental optical conductivity $\sigma_1(\omega)$ to the analytical response of a 1D Mott insulator derived in  Refs.~\cite{Essler2001excitons,mitrano2014pressure}. The reduced EHM optical conductivity $\omega\sigma_1(\omega)$ in the limit $U\gg t$ is:
\begin{equation}\label{eq:wsigma}
    \omega\sigma_1(\omega) = g_0t^2e^2\frac{1}{\hbar}\left\{\theta(V-2t)\pi\left[1-\left(\frac{2t}{V}\right)^2\right]\delta(\omega-\omega_\mathrm{CT})+\theta(4t-|\hbar\omega-U|)\frac{2t\sqrt{1-[\frac{\hbar\omega-U}{4t}]^2}}{V(\hbar\omega-\hbar\omega_\mathrm{CT})} \right\},
\end{equation}
where $e$ is the electron charge, $\theta(x)$ the Heaviside function, and $g_0=2.65$ the zero-momentum form factor accounting for the spin degrees of freedom. In the ``no-recoil" approximation, the optical response $\omega\sigma_1(\omega)$ is dominated by two $q=0$ transitions: (i) a Hubbard exciton located at $\hbar\omega_\mathrm{CT}=U-V-4t^2/V$~\cite{kim2008bound}, and (ii) a particle-hole continuum (PH) of unbound holon-doublon excitations contributing a broad peak centered around $U$ \cite{mitrano2014pressure}. We fit the reduced optical conductivity (normalized to its maximum) up to 3.5~eV to the expression in Eq.~\eqref{eq:wsigma}. We use a genetic algorithm to determine the optimal values of $U$, $V$, and $t$. The best-fit EHM parameters are $t=0.56$~eV, $U=3.96$~eV, and $V=1.17$~eV, which are used throughout the manuscript and are consistent with previous experimental findings~\cite{neudert1998manifestation}.

\section{NIR energy dependence of the THG renormalization}\label{sec:2Dmaps}\vspace{-3mm}

\begin{figure}[h]
\centering
\includegraphics[width=\textwidth]{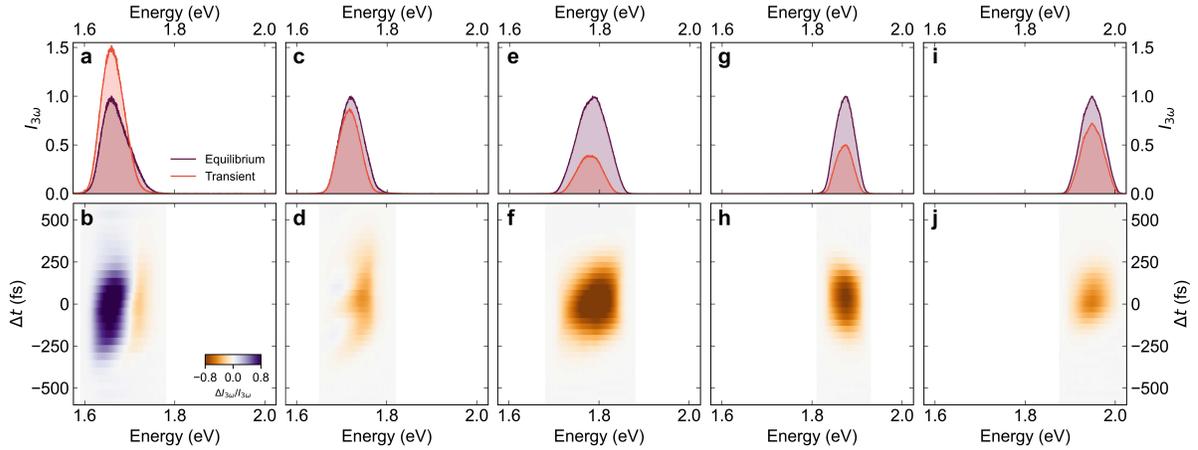}
\caption{\textbf{NIR energy dependence of the THG renormalization.} Time- and spectrally-resolved maps of the differential transient THG response. The driving frequency and the MIR peak electric field ($E_\mathrm{MIR}$) are: 0.56~eV and 1.0~MV/cm in \textbf{a} and \textbf{b}, 0.58~eV and 1.5~MV/cm in \textbf{c} and \textbf{d}, 0.59~eV and 1.8~MV/cm in \textbf{e} and \textbf{f}, 0.63~eV and 1.7~MV/cm in \textbf{g} and \textbf{h}, and 0.66~eV and 1.3~MV/cm in \textbf{i} and \textbf{j}. Each of the panels in the upper row shows one-dimensional spectra at maximum pump-probe overlap ($\Delta t=0$~fs) of the equilibrium (purple) and transient (orange) THG signal. The individual panels in the bottom row show the spectrally-resolved differential THG (normalized to the equilibrium signal) at different pump-probe delays ($\Delta t$).}\vspace{-12mm}
\label{fig:2Dmaps}
\end{figure}

In this section, we present the time- and spectrally-resolved transient THG response upon varying NIR probe energy. Figure~\ref{fig:2Dmaps} displays the differential THG intensity $\Delta I_{3\omega}(\omega)/I_{3\omega}(\omega)$ for five NIR energies between 0.56~eV and 0.66~eV. The MIR pump energy is held fixed at 0.12 eV (28~THz), and the MIR fluence for each dataset is reported in the caption. Note that while here the data is acquired at the maximum achievable field strength at each MIR wavelength, the wavelength-dependent data reported in the main text is measured while maintaining the same MIR field strength at each NIR wavelength.

\section{Dynamical THG renormalization of a light-driven extended Hubbard model}\label{sec:ED}
\begin{figure}[h]
\centering
\includegraphics[scale=0.725]{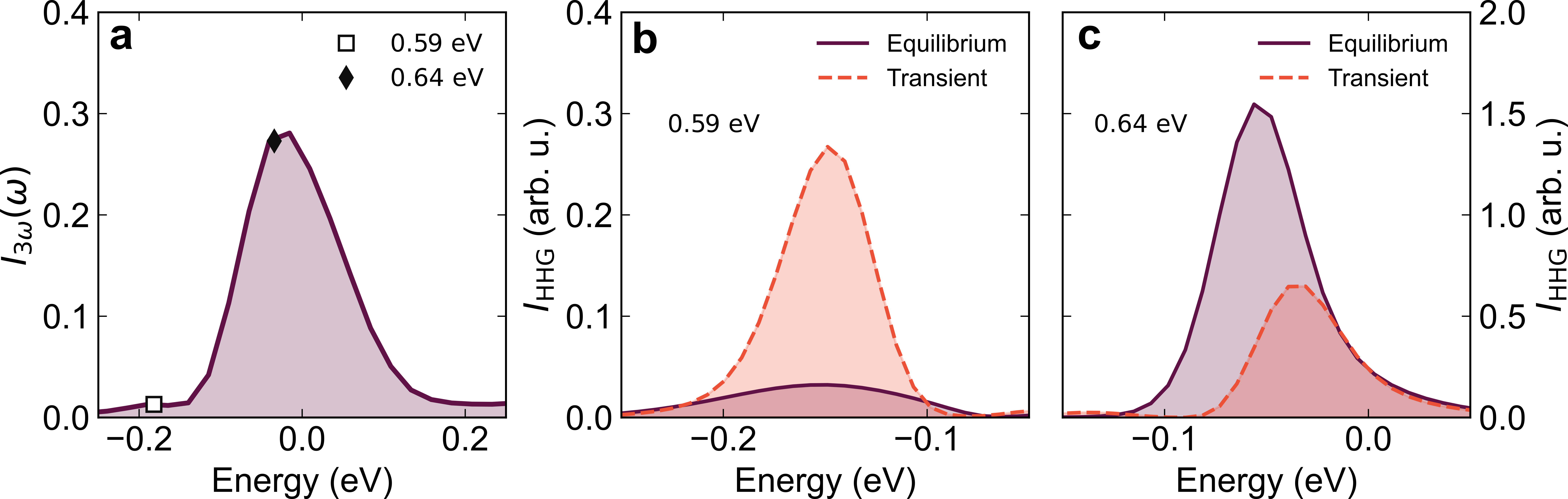}
\caption{\textbf{ED calculation of the high harmonic response.} Panel (a): Spectrally integrated static third-harmonic efficiency $I_{3\omega}^\mathrm{eq}$ at different NIR fundamental driving frequencies $\omega$, shown as a function of the photon energy $3\omega$ for a chain with $L=8$ sites. Markers indicate two selected driving frequencies: one resonant to the Mott gap (0.64~eV) and one red-detuned (0.59~eV) with respect to it. Panels (b) and (c) show the equilibrium and transient THG spectra for the two selected drivers, 0.59~eV (b) and 0.64~eV (c), respectively. All curves are referenced to the energy of the odd excitonic state.}\vspace{-3mm}
\label{fig:ED}
\end{figure}

\vspace{-3mm}
At equilibrium, $\mathrm{Sr_2CuO_3}$ is described by an EHM Hamiltonian at half filling
\begin{equation}
    \hat{H}_0 = -t \sum_{i,\sigma}^L\left(\hat{c}^{\dagger}_{i,\sigma}\hat{c}_{i+1,\sigma} + c. c.\right) + U\sum_i^L\hat{n}_{i,\uparrow}\hat{n}_{i,\downarrow} + V\sum_{i}^L\hat{n}_i\hat{n}_{i+1},
    \label{eq:ham_from_ed}
\end{equation}
where $\hat{c}^{\dagger}_{j,\sigma}$ ($\hat{c}_{j,\sigma}$) denotes the creation (annihilation) operator, $\hat{n}_{i,\sigma}=\hat{c}^{\dagger}_{j,\sigma}\hat{c}_{j,\sigma}$,  and $\hat{n}_{i}=\hat{n}_{i,\uparrow} + \hat{n}_{i,\downarrow}$. $U$ is the on-site Coulomb repulsion, while $V$ is a repulsive nearest-neighbour Coulomb interaction. From Sec.~\ref{sec:fit}, we fix the parameters to $U=7.05 t$ and $V=2.08 t$, with the hopping constant $t=0.56~\mathrm{eV}$. To simulate the exciton nonlinearity in presence of both NIR and MIR fields, we consider the combined electric field $\vec{E}(t) = \vec{E}_\mathrm{MIR}(t) + \vec{E}_\mathrm{NIR}(t)$, where each $\vec{E}_\Omega(t)$ is described by the following envelope:
\begin{equation}
    \vec{E}_\Omega(t) = 
    \begin{cases}
        E^0_\Omega \cos(\Omega t) \cos^2\left(\frac{\Omega t}{2 M_\mathrm{cyc}}\right), & \text{if}\ \--M_\mathrm{cyc}T_\Omega/2 < t < M_\mathrm{cyc}T_\Omega/2 \\
        0, & \text{otherwise}.
    \end{cases}
\end{equation}
$M_\mathrm{cyc}$ denotes the number of laser cycles and $T_\Omega = 2\pi/\Omega$ is the period of the NIR/MIR field. Fig.~\ref{fig:ED} shows calculations of the renormalized THG response of the Hubbard exciton on a $L=8$ sites chain with open boundary conditions. We fix  $M_\mathrm{cyc}=10$ for both fields, $E^0_\mathrm{MIR} = 1.5\ \mathrm{MV/cm}$, $E^0_\mathrm{NIR} = 2.5\ \mathrm{MV/cm}$, and $\Omega_\mathrm{MIR} = 136$ meV (32.9 THz). The NIR energy is varied between 0.54 eV (130 THz)and 0.79 eV (190 THz). We note that larger chains (up to $L=12$) consistently yield similar results, and the chain size only affects the position of the static THG peak due to the different energy level spacing upon varying $L$. We have verified that the results are robust against variations of the number of cycles, NIR/MIR field strengths, and MIR energy.

\par To obtain the nonlinear current, we introduce the coupling of the matter degrees of freedom with the vector potential of the laser field ($\vec{E}_\Omega(t) = -\partial_t \vec{A}_\Omega(t)$) via the Peierls substitution: $\cop_{i,\sigma}\rightarrow\cop_{i,\sigma}e^{-ia_0e\int\vec{A}_\Omega(t)\cdot\vec{r}_i\diffd \vec{r}}$, where $e$ is the elementary charge and $a_0$ is the lattice constant ($a_0=4.0$~\AA). We then numerically solve the associated time-dependent Schr\"odinger equation assuming a system initially in the ground state of $\hat{H}_0$ to obtain the time-evolved wavefunction $\ket{\Psi(t)}$. The time-dependent current is given by the expectation value of the current operator:
\begin{equation}
    \bra{\Psi(t)} \hat{j}(t) \ket{\Psi(t)} = -iea_0t\bra{\Psi(t)}\sum_{i,\sigma}^L\left(e^{-ia_0e\int\vec{A}_\Omega(t)\cdot\vec{r}_i d\vec{r}}\cop_{i,\sigma}^\dagger\cop_{i+1,\sigma} - \mathrm{H.c.}\right) \ket{\Psi(t)}.
    \label{eq:current_from_ed}
\end{equation}

\par Following Ref.~\cite{sundaram1990high}, we first calculate the dipole acceleration $a(t)=-\diffd \bra{\Psi(t)} \hat{j}(t) \ket{\Psi(t)} / \diffd t$ and obtain the harmonic spectrum through the Fourier transform of $a(t)$. Prior to that, the time-domain acceleration is multiplied by a Gaussian window function of the form: $e^{-t^2/\sigma_G^2}$ with $\sigma_G=53$~fs~\cite{silva2018high}.
Typical equilibrium and transient THG spectra are presented in Fig.~\ref{fig:ED}. In all ED THG calculations presented in this manuscript, we fix the NIR electric field at 2.5~MV/cm and the MIR at 1.5~MV/cm. We have confirmed that the dynamic effects observed in the ED calculations above remain valid in the infinite-size limit, free from finite-size effects. To this end, we performed infinite-size time-dependent density matrix renormalization group ($i$-trDMRG) calculations, which revealed the same qualitative features.

\section{Third-order Floquet susceptibility}\label{sec:chi3_Floquet}\vspace{-3mm}

\subsection{General Floquet analysis}\label{sec:Floq_generic}\vspace{-3mm}

The periodic driving of a many-body quantum state will lead to a dynamically renormalized nonlinearity which can be described through the Floquet formalism. Here, we explicitly derive the third-order susceptibility of a Floquet steady state of the EHM. Following Ref.~\cite{boyd2020nonlinear}, the induced polarization inside the material depends upon the expectation value of the dipole moment operator associated with the NIR-mediated transition:
\begin{equation}\label{eq:pdefBoyd}
    \left\langle \tilde{\bm{p}} \right\rangle \equiv \bra{\Psi(t)} \hat{\vec{\mu}} \ket{\Psi(t)}.
\end{equation}
In the above, $\ket{\Psi(t)}$ denotes the time-evolved wavefunction of the driven extended Hubbard chain under the action of the MIR pump. The Hamiltonian of this laser-driven system can be written as:
\begin{equation}
    \hat{H}(t) = \hat{H}_0 + \hat{H}_d(t),
\end{equation}
where $\hat{H}_0$ denotes the field-free EHM Hamiltonian and $\hat{H}_d(t) = -\muvec\cdot \vec{E}_\mathrm{MIR}(t)$ is the laser-matter interaction with the MIR field $\vec{E}_\mathrm{MIR}$ in the dipole approximation. If the laser field $\vec{E}_\mathrm{MIR}$ has a periodicity $T$ ($T=2\pi/\Omega$), then $\hat{H}_d(t)=\hat{H}_d(t+T)$ and we can expand $\Psi(t)$ in the basis of periodic Floquet functions $\left\{\Phi_\alpha\right\}$ as:
\begin{equation}
    \ket{\Psi(t)} = \sum_\alpha b_\alpha e^{-i\varepsilon_\alpha t/\hbar } \ket{\Phi_\alpha(t)}, 
\end{equation}
where 
\begin{equation}
    \ket{\Phi_\alpha(t+T)} = \ket{\Phi_\alpha(t)}
\end{equation}
and $\varepsilon_\alpha$ are the Floquet eigenenergies. The coefficients $b_\alpha$ are given by $b_\alpha = \langle \Phi_\alpha(t) | \Psi(t) \rangle $ and the Floquet eigenstates satisfy the orthonormality condition: $\langle \Phi_\alpha (t) | \Phi_\beta(t) \rangle= \delta_{\alpha\beta}$. $\Phi_\alpha(t)$ has the same periodicity as the MIR field and can be expanded in terms of the Floquet sidebands $\phi_\alpha^m$:
\begin{equation}
    \ket{\Phi_\alpha(t)} = \sum_m e^{-im\Omega t} \ket{\phi_\alpha^m}.
\end{equation}
The MIR-dressed chain is interrogated by the probe NIR field via:
\begin{equation}
    \hat{\mathcal{H}}(t) = \hat{H}_d(t) + \hat{V}(t),
\end{equation}
where 
\begin{equation}
    \hat{V}(t) = -\muvec \cdot \tilde{\vec{E}}_\mathrm{NIR}  = 
    -\muvec \cdot \sum_p \vec{E}_\mathrm{NIR}(\omega_p) e^{-i\omega_p t}
\end{equation}
is the perturbation corresponding to the NIR electric field $\vec{E}_\mathrm{NIR}(t)$. To calculate the induced polarization $\left\langle \tilde{\bm{p}} \right\rangle$, we follow Ref.~\cite{gu2018optical} and consider the time evolution operator:

\begin{equation}
    \hat{U}(t, t_0) = \mathcal{T}e^{-i/\hbar \int^t_{-\infty}\hat{\mathcal{H}}(t')\diffd t' } =  \hat{U}_d(t,t_0) \hat{S}(t, t_0),
\end{equation}
where $\mathcal{T}$ denotes time ordering. $\hat{U}_d(t,t_0)$ is the part corresponding to the MIR driving and can be calculated as:
\begin{equation}
    \hat{U}_d(t, t_0) = \sum_\alpha e^{-i \varepsilon_\alpha (t-t_0) /\hbar } \ket{\Phi_\alpha (t)}\bra{\Phi_\alpha(t_0)}.
\end{equation}
The operator $\hat{S}(t, t_0)$ incorporates the interaction with the probe light and is related to the perturbation $\hat{V}(t)$ in the interaction picture:
\begin{equation}
    i\hbar \partial_t \hat{S}(t, t_0) = \hat{V}_I(t) \hat{S}(t, t_0),
\end{equation}
where 
\begin{equation}
    \hat{V}_I(t) = \hat{U}_d^\dagger (t, t_0) \hat{V}(t) \hat{U}_d(t,t_0).
\end{equation}
$\hat{S}(t,t_0)$ can be represented by a Dyson series expansion:
\begin{equation}\label{eq:Dyson}
    \hat{S}(t, t_0) = \mathbb{I} + \sum_{n=1}^\infty \left(\frac{-i}{\hbar}\right)^n \int^t \diffd t_1 \int^{t_1} \diffd t_2 \ldots \int^{t_{n-1}}\diffd t_n 
    \hat{V}_I(t_1) \hat{V}_I(t_2) \ldots \hat{V}_I(t_n).
\end{equation}
$\hat{U}(t,t_0)$ can be used to obtain the time-evolved $ \ket{\Psi(t)} $ from the wavefunction at time $t_0$:
\begin{equation}\label{eq:PsiTU}
    \ket{\Psi(t)} = \hat{U}(t,t_0) \ket{\Psi(t_0)} \equiv \hat{U}(t,t_0)\ket{\Psi_g},
\end{equation}
where 
\begin{equation}
    \ket{\Psi_g} = \sum_\alpha b_\alpha e^{-i\varepsilon_\alpha t_0/ \hbar } \ket{\Phi_\alpha (t_0)}
\end{equation}
is the MIR-dressed wavefunction prior to the interaction with the probe field. We now insert Eq.~\eqref{eq:PsiTU} in Eq.~\eqref{eq:pdefBoyd}:
\begin{equation}
    \left\langle \tilde{\bm{p}} \right\rangle = \bra{\Psi_g} \hat{S}^\dagger(t, t_0) \hat{U}_d^\dagger (t, t_0)  \muvec \hat{U}_d(t, t_0) \hat{S}(t, t_0) \ket{\Psi_g}.
\end{equation}
and expand $\hat{S}(t, t_0)$ up to the third order:
\begin{equation}
\hat{S}(t, t_0) = \mathbb{I} + \hat{S}^{(1)}(t, t_0) + \hat{S}^{(1)}(t, t_0) + \hat{S}^{(3)}(t, t_0).
\end{equation}
The third-order contribution to $\left\langle \tilde{\bm{p}} \right\rangle$ can be written as:
\begin{eqnarray}
    \left\langle \tilde{\bm{p}}^{(3)} \right\rangle & = & 
    \bra{\Psi_g}  \hat{U}_d^\dagger (t, t_0) \, \muvec \,\, \hat{U}_d(t, t_0)\, \hat{S}^{(3)}(t, t_0) \ket{\Psi_g} +\nonumber  \\ 
    & + & 
    \bra{\Psi_g} \hat{S}^{(1)\dagger}(t, t_0) \, \hat{U}_d^\dagger (t, t_0) \, \muvec \,\, \hat{U}_d(t, t_0) \, \hat{S}^{(2)}(t, t_0) \ket{\Psi_g} +\nonumber  \\
    & + & 
    \bra{\Psi_g} \hat{S}^{(2)\dagger}(t, t_0) \, \hat{U}_d^\dagger (t, t_0) \, \muvec \,\, \hat{U}_d(t, t_0) \,\hat{S}^{(1)}(t, t_0) \ket{\Psi_g} +\nonumber  \\
    & + & 
    \bra{\Psi_g} \hat{S}^{(3)\dagger}(t, t_0) \, \hat{U}_d^\dagger (t, t_0) \,  \muvec \,\, \hat{U}_d(t, t_0) \ket{\Psi_g},
\end{eqnarray}
where:
\begin{equation}
    \hat{S}^{(n)}(t,t_0) = \left(\frac{-i}{\hbar}\right)^n \int^t \diffd t_1 \ldots \int^{t_{n-1}} \diffd t_n \hat{V}_I (t_1) \ldots \hat{V}_I(t_n).
\end{equation}
After a series of substitutions, we obtain the third-order susceptibility for a Floquet system in the frequency domain:
\begin{eqnarray} \label{eq: THG_floquet}
    \chi^{(3)}(-\omega_\sigma; \omega_p, \omega_q, \omega_r) & = & \frac{1}{\hbar^3}\sum_{\alpha\delta}\sum_{pqr}\sum_{\gamma\mu\nu}\sum_{m_1l_1}\sum_{m_2l_2}\sum_{m_3l_3}\sum_{m_4l_4} b_\alpha^\ast b_\delta \exp\left\{ i\left[ \Omega_{\alpha\delta} - \omega_r - \omega_q - \omega_p \right] t \right\} \times \nonumber \\
    & \times & \muvec_{\alpha\gamma}^{m_1,m_1+l_1} \muvec_{\gamma\mu}^{m_2,m_2+l_2} \muvec_{\mu\nu}^{m_3,m_3+l_3}\muvec_{\nu\delta}^{m_4,m_4+l_4} \times \nonumber\\
     & \times & \left(  \frac{1}{( \Omega_{\gamma\delta} - \Omega^{l_2l_3l_4}  - \omega_r - \omega_q - \omega_p)(  \Omega_{\mu\delta} - \Omega^{l_3l_4} - \omega_q - \omega_p)(  \Omega_{\nu\delta} - \Omega^{l_4} -  \omega_p)}  + \right. \nonumber \\
    & + & \frac{1}{( \Omega_{\gamma\alpha}^\ast - \Omega^{l_2l_3l_4}  + \omega_r )(  \Omega_{\mu\delta} - \Omega^{l_3l_4} - \omega_q - \omega_p)(  \Omega_{\nu\delta} - \Omega^{l_4} -  \omega_p)}  +  \\
     & + & \frac{1}{( \Omega_{\gamma\alpha}^\ast - \Omega^{l_2l_3l_4}  + \omega_r )(  \Omega_{\mu\alpha}^\ast - \Omega^{l_3l_4} + \omega_r + \omega_q)(  \Omega_{\nu\delta} - \Omega^{l_4} -  \omega_p)}  + \nonumber \\
     & +  & \left. \frac{1}{( \Omega_{\gamma\alpha}^\ast - \Omega^{l_2l_3l_4}  + \omega_r )(  \Omega_{\mu\alpha}^\ast - \Omega^{l_3l_4} + \omega_r + \omega_q)(  \Omega_{\nu\alpha}^\ast - \Omega^{l_4} + \omega_r + \omega_q +  \omega_p)}\right),\nonumber
\end{eqnarray}
where we have introduced the short-hand notations $\Omega^{l_1l_2\ldots l_k} = (l_1+l_2+\ldots + l_k)\Omega$ and $\Omega_{\alpha\beta} = \tfrac{1}{\hbar}(\varepsilon_\alpha - \varepsilon_\beta) - i\gamma$. Here, $\gamma$ is a phenomenological broadening factor which we set to $\hbar\gamma=0.06 t$. Note that for third harmonic generation ($\chi^{(3)}(-3\omega; \omega, \omega, \omega)$), the above sum is restricted to the condition $l_1+l_2+l_3+l_4=0$. The dressed dipole moment connecting individual Floquet sidebands is defined as $\muvec_{\alpha\gamma}(t) = \bra{\Phi_\alpha(t)}\muvec\ket{\Phi_\gamma(t)} = \sum_{mn}e^{i(m-n)\Omega t}\bra{\phi_\alpha^m}\muvec \ket{\phi_\gamma^n} = \sum_{mn}e^{i(m-n)\Omega t}\muvec_{\alpha\gamma}^{mn}$.\vspace{-3mm}

\subsection{Three-state model}\label{sec:3S}\vspace{-3mm}

In this section, we outline the phenomenological three-level system (3LS) model describing how the off-resonant MIR driving renormalizes the THG spectrum. Within this model, we consider only the ground state ($\ket{0}$) as well as the pair of even ($\ket{g}$; \textit{gerade}) and odd ($\ket{u}$; \textit{ungerade}) Hubbard excitons. Since the pump (0.12~eV) is far off-resonance with the charge gap ($\sim1.8$~eV), we neglect the Floquet dressing of the ground state by the drive. Further, we take the even and odd Mott excitons to be exactly degenerate with energy $\epsilon_{ex}$ and denote the dipole matrix element between even and odd exciton states along the chain direction by $\mu_{ug}$. When driven by a \textit{continuous-wave} (cw) oscillating electric field, the 3LS Hamiltonian ($\hat{H}_\mathrm{3LS}$) takes the following form:
\begin{equation}
\hat{H}_\mathrm{3LS}(t) = \epsilon_{ex} \ket{u} \bra{u} + \epsilon_{ex} \ket{g} \bra{g} + \mu_{ug} \mathcal{E} \left( \ket{u} \bra{g} + \ket{g} \bra{u} \right)  \cos{\Omega t},
\end{equation}
where $\mathcal{E}\equiv E_\mathrm{MIR}^0$ is the pump field polarized along the chain direction and $\Omega\equiv\Omega_\mathrm{MIR}$ denotes its frequency. In other words, for simplicity we have set $\mathcal{E}\equiv E_\mathrm{MIR}^0$ and $\Omega\equiv\Omega_\mathrm{MIR}$. 

In the Fourier space, the above Hamiltonian becomes:
\begin{equation}\label{eq:H3LSF}
\begin{split}
\hat{H}_\mathrm{3LS}^F = & \epsilon_{ex} \left(\ket{u} \bra{u} + \ket{g} \bra{g}\right) \otimes \mathbb{I}_m
- \sum_m m \mathbb{I} \otimes \ket{m} \bra{m} \\
& + \sum_m \frac{\mu_{ug} \mathcal{E}}{2} 
\left(
\ket{u} \bra{g} +  \ket{g} \bra{u})
\right)
\otimes
\left( \ket{m+1} \bra{m} +  \ket{m} \bra{m+1}
\right).
\end{split}
\end{equation}
While the \textit{static} inversion symmetry is broken on a subcycle scale, the Hamiltonian $\hat{H}_\mathrm{3LS}^F$ is characterized by a \textit{dynamical} inversion symmetry. In more detail, $\hat{H}_\mathrm{3LS}^F$ is invariant under time translation by half a period followed by inversion, that is, $\hat{H}_\mathrm{3LS}^F(t) = \hat{\Pi} \hat{H}_\mathrm{3LS}^F(t + \frac{T}{2}) \hat{\Pi}$, where $T$ is the MIR period and $\hat{\Pi}$ is the inversion operator. Similarly to the case of static inversion symmetry, we can choose the eigenstates to be either even or odd under the dynamical symmetry operation. Based on these considerations, we write down the following \textit{ansatz} for the dressed exciton eigenstates: 
\begin{equation}\label{eq:u_tilde}
\ket{\Tilde{u}} =
\sum_{m}
\phi^u_{2m}  \ket{u} \otimes \ket{2m} + 
\phi^u_{2m+1} \ket{g} \otimes \ket{2m+1}
\end{equation}
and:
\begin{equation}\label{eq:g_tilde}
\ket{\Tilde{g}} =
\sum_{m}
\phi^g_{2m} \ket{g} \otimes \ket{2m} + 
\phi^g_{2m+1} \ket{u} \otimes \ket{2m+1}.
\end{equation}
By inserting these eigenstates into the Fourier-transformed Hamiltonian~\eqref{eq:H3LSF}, we find that the exciton energy does not change under intense pump driving and that $\phi^{u,g}$ satisfy the recursion relationship of Bessel functions. Based on these considerations, we derive the following expression for the dressed states:
\begin{equation} \label{eq:Ex_u}
\ket{\Tilde{u}} =
\sum_{m}
J_{2m} \left( \frac{\mu_{ug}\mathcal{E}}{\hbar\Omega} \right) \ket{u} \otimes \ket{2m} + 
J_{2m+1}\left( \frac{\mu_{ug}\mathcal{E}}{\hbar\Omega} \right) \ket{g} \otimes \ket{2m+1}
\end{equation}
and:
\begin{equation} \label{eq:Ex_g}
\ket{\Tilde{g}} =
\sum_{m}
J_{2m}\left( \frac{\mu_{ug}\mathcal{E}}{\hbar\Omega} \right) \ket{g} \otimes \ket{2m} + 
J_{2m+1}\left( \frac{\mu_{ug}\mathcal{E}}{\hbar\Omega} \right) \ket{u} \otimes \ket{2m+1}.
\end{equation}
With these states, we calculate the THG intensity using Eq.~\eqref{eq: THG_floquet} with a focus on the contribution from the term corresponding to the three-photon resonance:
\begin{equation} \label{eq:THG_3S}
\chi^{(3)}(-3 \omega; \omega, \omega, \omega) 
~\propto~
\sum_n
\frac{\left[J_n \left(\frac{\mu_{ug} \mathcal{E}}{\hbar\Omega} \right)\right]^2 \mu_{ug}^2 \mu_{0u}^2 }
{ (\epsilon_{ex}/\hbar + n \Omega - 3 \omega - i \gamma)
(\epsilon_{ex}/\hbar + n \Omega - 2 \omega - i \gamma)
(\epsilon_{ex}/\hbar + n \Omega - \omega - i \gamma) } ,
\end{equation} 
where $\mu_{u0} \equiv \left \langle u | \hat{\mu}_b | 0 \right \rangle$ is the dipole matrix element between the equilibrium ground state and the undressed odd exciton.\vspace{-3mm}

\subsubsection{Rabi oscillations}\label{sec:Rabi}

\begin{figure}[h]
\centering
\includegraphics[width=0.7\textwidth]{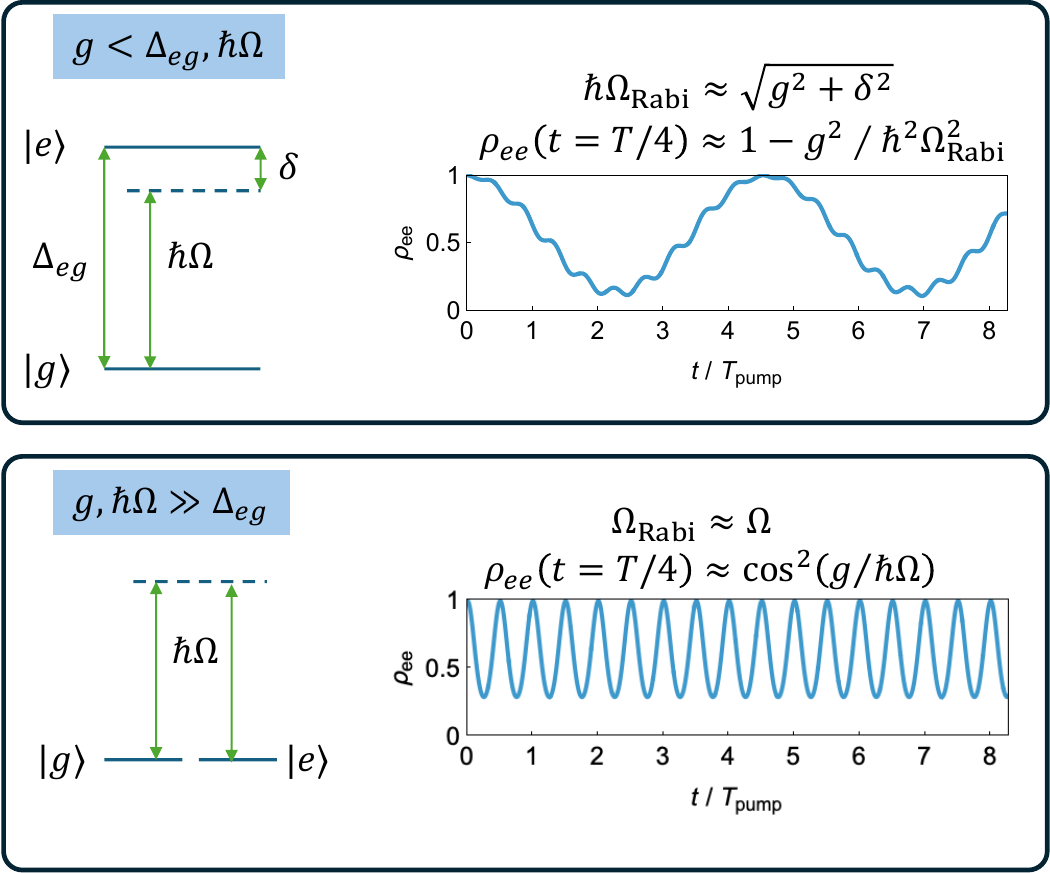}
\caption{\textbf{Rabi oscillations in the strong-coupling limit for degenerate states $|e\rangle$, $|g\rangle$.} Comparison between the conventional quantum-optical Rabi problem (top panel) and the strong-coupling Rabi problem for a degenerate two-level system (bottom panel). In the quantum-optical Rabi problem, characterized by a near-resonant driving field $\hbar\Omega \sim \Delta_{eg}$ and weak light-matter coupling $g$, the $|e\rangle$ state population oscillates with a Rabi frequency $\Omega_{\rm Rabi} \approx g$. In contrast, in the Rabi problem relevant for driven Hubbard excitons, the light-matter coupling strength $g$ becomes comparable to the pump frequency $\hbar \Omega$, which in turn greatly exceeds a vanishing exciton splitting $\Delta_{eg} \to 0$. Here, the Rabi frequency is pinned to the driving frequency $\Omega_{\rm Rabi} \approx \Omega$, while the Rabi amplitude is determined by $g/\hbar \Omega$.}
\label{fig:rabi}
\end{figure}

The Floquet solution for the two-level system comprised of degenerate even and odd Hubbard excitons can be understood as a strong-coupling limit of the quantum-optical Rabi Hamiltonian. In atomic settings, the light-matter coupling $g = \vec{\mu}_{eg} \cdot \vec{E}$  for dipole-active transitions between two ground ($|g\rangle$) and excited states ($|e\rangle$) is much smaller than their energy difference $\Delta_{eg}$. Consequently, a rotating-wave approximation is well-justified near resonance $\hbar\Omega \approx \Delta_{eg}$ and the atomic state occupation oscillates with a Rabi frequency $\Omega_{\rm Rabi} \approx \sqrt{g^2 + (\hbar \Omega - \Delta_{eg})^2}$. In contrast, the light-matter coupling $g$ for driven Hubbard excitons considered in the main text becomes comparable to the pump frequency $\Omega$, which in turn is highly off-resonant from the dipole transition between the nearly-degenerate exciton states. In this limit, described in the main text, the Rabi frequency is pinned to the pump frequency $\Omega_{\rm Rabi} \approx \Omega$ while the Rabi amplitude is set by the light-matter coupling strength $g/\hbar\Omega$.

\subsubsection{Extraction of the Bloch rotation angle}\vspace{-3mm}

\begin{figure}[h]
\centering
\includegraphics[width=\textwidth]{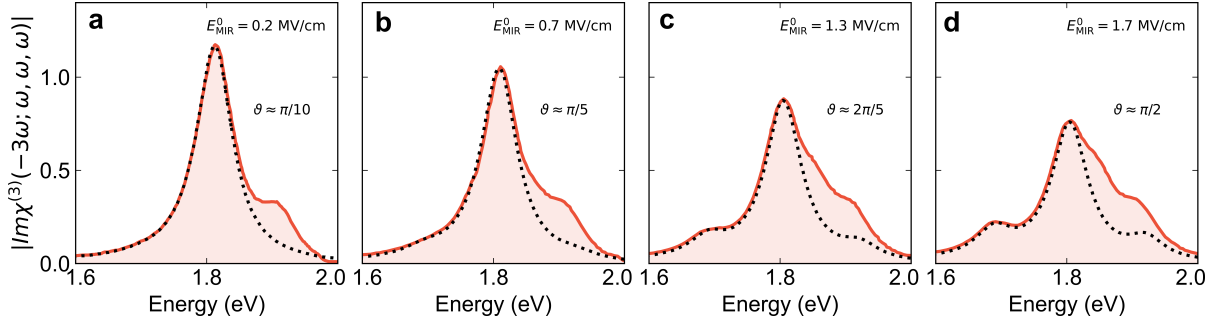}
\caption{\textbf{Extraction of the Bloch rotation angle from the driven HD model.} Comparison between imaginary third-order suceptibility obtained by solving the driven HD model (orange lines) and that predicted by the driven 3LS model with adjusted parameters (black dashed lines). The individual panels ((a)-(d)) correspond to different MIR pump strengths. The extracted Bloch angle rotation is indicated in each panel. }\vspace{-7mm}
\label{fig:fits}
\end{figure}

As discussed in greater detail in Sec.~\ref{sec:HD_3LS_rel}, the presence of the holon-doublon (HD) continuum renormalizes the bare exciton response expected from the three-state model. To extract the rotation angle from the driven holon-doublon third-harmonic spectrum, we follow the procedure outlined below. First, the exciton energy $\epsilon_{ex}$ is determined by identifying the energy of the main peak. Second, the effective dipole matrix element $\mu_{ug}$ is obtained by comparing the relative heights of the main peak and the side peak in the THG spectrum with the prediction of Eq.~\eqref{eq:THG_3S}. Lastly, since $\mu_{u0}$ is an overall prefactor in Eq.~\eqref{eq:THG_3S}, its value is chosen to ensure that the height of the main peak in the three-state THG matches that of the holon-doublon THG (see Fig.~\ref{fig:fits}). The rotation angle depends solely on the effective value of $\mu_{ug}$. Once this value is determined, the rotation on the Bloch sphere can be calculated using the exact solutions of the dressed exciton wavefunctions in Eqs.~\eqref{eq:Ex_g} and~\eqref{eq:Ex_u}.

\par The presence of the continuum complicates the intracycle dynamics of the exciton wavefunction compared to a two-level Bloch-sphere system, due to mixing with unbound particle-hole states. Rather than a simple rotation around the Bloch sphere, the exciton evolves within a higher-dimensional Bloch sphere. This motion parallels the two-level system, but instead of oscillating between even and odd states, the exciton transitions between the even and odd parity sectors of the Hilbert space. Consequently, the cycle-averaged parity deviates from $\pm 1$, reflecting the state trajectory around the higher-dimensional Bloch sphere, a many-body analog of the rotation angle. As discussed in Sec.~\ref{sec:HD_3LS_rel}, a change of basis can reduce this dynamics to an effective two-level system with renormalized parameters. Thus, the cycle-averaged parity can still be interpreted as an effective motion on a Bloch sphere.  \vspace{-3mm}

\subsubsection{Simplified Bloch angle estimation}\label{sec:Bloch-simple}
The Bloch spheres in the insets of Fig.~3 of the main text have been obtained using a simpler procedure based on a three-level system comprised of a ground state and two degenerate excited states, the latter being Floquet-dressed by the MIR field. Using the \textit{ansatz} of Eqs.~\ref{eq:u_tilde} and~\ref{eq:g_tilde}, we have:
\begin{eqnarray}
    \ket{\tilde{u}} & = & \sum_m e^{2im\Omega t} J_{2m}\left(\frac{\mu_{ug}\mathcal{E}}{\hbar\Omega}\right)\ket{u} + e^{(2m+1)i\Omega t} J_{2m+1}\left(\frac{\mu_{ug}\mathcal{E}}{\hbar\Omega}\right)\ket{g} \nonumber \\
    & = & \cos\frac{\vartheta}{2}\ket{u} + \sin\frac{\vartheta}{2}\ket{g}\\
    \ket{\tilde{g}} & =& \sum_m e^{2mi\Omega t} J_{2m}\left(\frac{\mu_{ug}\mathcal{E}}{\hbar\Omega}\right)\ket{g} + e^{(2m+1)i\Omega t} J_{2m+1}\left(\frac{\mu_{ug}\mathcal{E}}{\hbar\Omega}\right)\ket{u} \nonumber \\
    & = & \cos\frac{\vartheta}{2}\ket{g} + \sin\frac{\vartheta}{2}\ket{u}.
\end{eqnarray}
The Bloch angle is given by $\vartheta=2\arccos\left[ \sum_m J_{2m}\left(\frac{\mu_{ug}\mathcal{E}}{\hbar\Omega}\right)e^{2mi\Omega t}\right])$. The Bloch angle in the middle inset of Fig.~3 of the main text amounts to 121.66 deg. We used a MIR energy of 125~meV and the experimentally determined dipole moment of 7.37~\AA~\cite{Kishida2000gigantic}. We also assumed a sinusoidal field form with $E_\mathrm{MIR}=1.8$~\mbox{MV/cm} and evaluated the Bloch angle at the maximum of the field, that is, at $t=T_\mathrm{MIR}/4$.\vspace{-3mm}

\subsection{Holon-Doublon model}\label{sec:HD}\vspace{-3mm}

The large optical nonlinearity of Mott insulating chains has traditionally been rationalized by modeling photoexcited states within a holon-doublon (HD) model \cite{mizuno2000nonlinear,kishida2001large}. The spin degrees of freedom are effectively decoupled from the problem because of the inherent spin-charge separation of 1D systems. In the HD model, the ground state is a vacuum where all sites are singly-occupied and the excitations are doubly-occupied (doublons) and unoccupied (empty) sites (holons). Following \cite{mizuno2000nonlinear}, we define the equilibrium HD Hamiltonian as:
\begin{equation}\label{eq:HD}
    \hat{H}_{hd} = - t \sum_i \left ( \hat{h}^{\dagger}_{i+1} \hat{h}_i + \hat{d}^{\dagger}_{i+1} \hat{d}_i + h.c. \right)
    - V \sum_{\left < i,j
    \right > } \hat{h}^{\dagger}_i \hat{h}_i \hat{d}^{\dagger}_j \hat{d}_j + \frac{U}{2} \sum_i \left( 
    \hat{h}^{\dagger}_i \hat{h}_i 
    + 
    \hat{d}^{\dagger}_i \hat{d}_i
    \right), 
\end{equation}
alongside the condition that $\hat{h}^{\dagger}_i \hat{h}_i + \hat{d}^{\dagger}_i \hat{d}_i \leq 1$ for all sites $i$. Here, $\hat{h}^{\dagger}_i$ ($\hat{d}^{\dagger}_i$) is the operator that creates a holon (doublon) on site $i$. Note that we have made a gauge transformation, so that the hopping of both holons and doublons has the same sign. As we focus on the features near the excitonic peak, we constrain our analysis to the sector where $ \sum_i \hat{h}^{\dagger}_i \hat{h}_i = \sum_i \hat{d}^{\dagger}_i \hat{d}_i = 1$. The HD model allows us to access larger system sizes than the extended Hubbard model, thus better capturing the effects from the continuum of unbound holon-doublon excitations.

\par To include the effects of the Floquet dressing, we couple the holon-doublon Hamiltonian to the electric field in the dipole gauge. This leads to:

\begin{equation}\label{eq:HD_t}
\hat{H}_{hd}^F(t) = \hat{H}_{hd} + \hat{P}_\mathrm{dip}\mathcal{E}(t),
\end{equation}
where, as before, $\mathcal{E}(t) = \mathcal{E} \cos{\Omega t}$ is the MIR electric field ($\Omega$ is the angular frequency of the MIR pump) and $\hat{P}_\mathrm{dip} = \sum_{i} ia (\hat{h}^{\dagger}_i \hat{h}_i - \hat{d}^{\dagger}_i \hat{d}_i)$ is the dipole operator along the chain ($a=ea_0$ is the dipole moment corresponding to an adjacent holon-doblon pair). We then use the extended-space formalism to find the Floquet eigenstates, including $40$ sidebands to reach convergence. We then use formula~\eqref{eq: THG_floquet} to calculate the THG spectrum of the system. Since the pump is far below resonance with respect to the charge gap, the equilibrium ground state is not appreciably dressed by the drive. As such, when applying Eq.~\eqref{eq: THG_floquet}, we assume that the system is in its ground state before the probe photoexcitation process.

\par Since the holon-doublon model is an effective description of the photoexcited states of the Hubbard model, the Hamiltonian parameters ($t$, $U$, $V$) are slightly renormalized from their corresponding values in the extended Hubbard model. The physics of the holon-doublon model is essentially controlled by $V/t$ as $U$ acts only as an overall chemical potential for the photoexcited states. To determine the appropriate parameter values, we use the methods described above to simulate the THG at different pump field strengths for various values of $V/t$ and broadening factors $\hbar\gamma/t$ to determine the set of parameters that agree best with the experimental data. To compare with the experiment, we focus on the suppression of the main peak intensity. Upon varying $V/t$, we set $U$ to a value such that the energy of the excitonic peak is fixed at $U - V - \frac{4t^2}{V} = 1.8$~eV. Lastly, because the dipole elements of the holon-doublon model are inherently arbitrary up to an overall scaling factor - and to account for any reflection, pump envelope shape effects, etc. - we include a rescaling factor between the simulated pump field strength and the experimental one ($E_{\mathrm{MIR}}^{\mathrm{exp}}/E_{\mathrm{MIR}}^{\mathrm{th}}=1.85$). In the numerical results reported in the manuscript, we use $t = 0.56$~eV, $V/t = 2.4$, a broadening factor of $\hbar \gamma/t = 0.06$, and a chain length of 16 sites. \vspace{-3mm}

\subsubsection{Relation to the three-state model}\label{sec:HD_3LS_rel}\vspace{-3mm}
The three-state model can be recovered by restricting the analysis to the Hubbard excitons. In this section, we show that one can take the Floquet Hamiltonian for the driven holon-doublon model and integrate out the holon-doublon continuum to arrive at a model that describes driven Hubbard excitons.

\par We choose to work in an eigenbasis of the dipole operator. We label each state as $\left | i, \mu\right \rangle \otimes \left | m \right \rangle$ where $i$ is the site of the holon operator, $\mu$ denotes the dipole element of the state, and $m$ is the Floquet replica index. The states with $\mu = \pm 1$ are excitons, whereas states with $|\mu| > 1$ correspond to unbound holon-doublon excitations.

\par If the hopping is turned off ($t = 0$), then $\hat{H}_{hd}^F$ can be solved exactly. Since each state is an eigenstate of $\hat{P}_\mathrm{dip}$, $\left | i, \mu\right \rangle \otimes \left | m \right \rangle$ only mixes with states having the same values of $i$ and $\mu$. Therefore, the Hamiltonian is:
\begin{equation}
\begin{pmatrix}
 ... &  &  &  & \\
 & E + \hbar\Omega & \frac{\mathcal{E} a\mu}{2} & 0 & \\
 & \frac{\mathcal{E} a\mu}{2} & E & \frac{\mathcal{E} a\mu}{2}& \\
 & 0 &  \frac{\mathcal{E}a \mu}{2} & E - \hbar \Omega & \\
  &  &  & & ... \\
\end{pmatrix},
\end{equation}
where $E$ is the equilibrium energy of the state ($E = U - V$ for the excitons and $E=U$ for the unbound states) and $\mathcal{E}$ is the field strength. Note that this Hamiltonian is just the three-state model, so we can diagonalize it exactly. Thus, the eigenstates of the above Hamiltonian become:
\begin{equation}
\left | i, \mu, m \right \rangle =
\sum_n J_{n-m} \left (
\frac{\mathcal{E} a\mu}{\hbar \Omega}
\right ) \left | i, \mu\right \rangle \otimes \left | n \right \rangle.
\end{equation}
For $t \neq 0$ but $t \ll V$, we can perform a Schrieffer-Wolff transformation to build an effective Hamilton that only describes the excitons and their Floquet replicas. The basic picture is that a finite $t$ lets the exciton break up into an unbound holon-doublon pair in the continuum and then recombine and come back to the exciton band. We wish to build an effective Hamiltonian which integrates out those processes. It is useful to define $\Tilde{J}_m \left (x \right ) = \sum_n J_{n-m}(2x) J_n(x)$ and
\begin{equation}
\Tilde{t}_{m,\mu} = \frac{1}{2} t^2 \sum_{l}
\Tilde{J}_{l-m}\left ( \frac{\mathcal{E}a \mu}{\hbar\Omega} \right )
\Tilde{J}_l\left ( \frac{\mathcal{E} a\mu}{\hbar\Omega} \right ) 
\left ( \frac{1}{(l -m)\hbar \Omega + V }
+ 
\frac{1}{V + l\hbar  \Omega}
\right )  .
\end{equation}
With these definitions, the effective Hamiltonian for the excitons becomes:
\begin{equation}
H_{eff} = U- V -
\sum_{\left \langle i,j \right \rangle}
\sum_{m,n}
\sum_{\mu = \pm 1}
\Tilde{t}_{m-n,\mu}
\left | i, \mu, m \right \rangle
\left \langle j, \mu, n \right |
- \sum_{i,m, n} \sum_{\mu = \pm 1}
(2 \Tilde{t}_{m-n,\mu} - m\hbar \Omega \delta_{mn}) \left | i, \mu, m \right \rangle
\left \langle i, \mu, n \right | .
\end{equation}
This Hamiltonian describes how excitons of each orientation (i.e, $\mu=-1$ for the ``holon-doublon'' configuration and $\mu=1$ for the ``doublon-holon'' one) can hop between sites and also hop to different Floquet sidebands. Because this model is translationally invariant, it is natural to perform a Fourier transform. This yields:
\begin{equation}
\hat{H}_{eff}(k) = U- V - 2 \cos{k}
\sum_{m,n}
\sum_{\mu = \pm 1}
\Tilde{t}_{m-n,\mu}
\left | k, \mu, m \right \rangle
\left \langle k, \mu, n \right |
- \sum_{i,m} \sum_{\mu = \pm 1}
(2 \Tilde{t}_{m-n,\mu} - m\hbar  \Omega \delta_{mn}) \left | k, \mu, m \right \rangle
\left \langle k, \mu, n \right |  .
\end{equation}
Since the ground state has $k=0$, we can work in the momentum sector where the Hamiltonian is:
\begin{equation}
\hat{H}_{eff}(k = 0) = U- V 
- \sum_{i,m, n} \sum_{\mu = \pm 1}
(4 \Tilde{t}_{m-n,\mu} - m\hbar  \Omega \delta_{mn}) \left | k = 0, \mu, m \right \rangle
\left \langle k = 0,  \mu, n \right | .
\end{equation}
Note that there are only two exciton states in this sector and, moreover, by taking even and odd superposition, we recover excitonic states which are even/odd under parity. When we consider these two excitons and the ground state, we recover the three-state model. Converting back into a time dependent Hamiltonian, we arrive at:
\begin{equation}
\hat{H}_{eff}(t) = 
\begin{pmatrix}
0 & 0 & 0 \\
0 & \Tilde{E}_{ex} + \Tilde{X}_e(t) & \Tilde{X}_o(t) \\
0 & \Tilde{X}_o(t) & \Tilde{E}_{ex} + \Tilde{X}_e(t) 
\end{pmatrix},
\end{equation}
where
\begin{equation}
\Tilde{E}_{ex} = U - V - 4 \Tilde{t}_{0,+1}
\ \ \ \ \ \
\Tilde{X}_e(t) = -8 \sum_{i=1}^{\infty} \Tilde{t}_{2i,+1} \cos\left(2i \Omega t\right),
\ \ \ \ \ \
\Tilde{X}_o(t) = - 8 \sum_{i=1}^{\infty} \Tilde{t}_{2i-1,+1} \cos\left((2i-1) \Omega t\right).
\end{equation}
In the above, we have used that $\Tilde{t}_{m,-1} = (-1)^m \Tilde{t}_{m,+1}$ in order to combine terms. This is simply the original three-state model, but now there additional couplings which come from integrating out the continuum modes. Note that the expansion parameter for this Hamiltonian is $t/V$. In the experiment and the numerical calculations, $t/V \sim \frac{1}{2}$ which is outside of this perturbative regime. In light of this, the predictions from this effective Hamiltonian are useful in so far as they allow for a qualitative picture of how the pump renormalization process occurs.

\par The most important insight from this effective Hamiltonian is related to exciton energy. $\Tilde{t}_{0}$ corresponds to the hopping in the same Floquet sideband and contributes to the overall energy of the exciton. For small fields, we can expand this as:
\begin{equation}
\Tilde{t}_{0} = \frac{t^2}{V} + \frac{1}{2} \frac{a^2 t^2}{V (V- \hbar\Omega) (V + \hbar \Omega)} \mathcal{E}^2 + ... ,
\end{equation}
where $a$ is the dipole moment of the bare exciton. The first term is present at equilibrium while the second term originates from pump-induced processes which increase the hopping amplitude, redshifting the excitonic band edge.\vspace{-3mm}

\subsubsection{Coherent nature of the quantum state manipulation}\label{sec:coherence}\vspace{-3mm}

In this section, we examine the coherence of the Floquet control protocol by analyzing the role of the exciton lifetime in its Floquet dressing. The exciton lifetime can be estimated from photodoping experiments that generate holon-doublon pairs across the Mott gap, yielding a value of approximately 2 ps \cite{ogasawara2000ultrafast}, corresponding to more than 30 MIR cycles. Within the three-state model, we incorporate decoherence via a Lindblad operator, $J = \sqrt{\nu} \ket{0}\bra{u}$, representing the radiative recombination of the odd exciton to the ground state \cite{lindblad1976generators, gorini1976completely}. Here, $\nu/\hbar$ denotes the inverse exciton lifetime. Radiative decay of the even exciton is forbidden by dipole selection rules.

We simulate the pump-probe experiment using a three-level system with Lindblad recombination to account for the finite exciton lifetime. The exciton energy is fixed at $\epsilon_{ex} = 1.8$ eV, and the dipole matrix element is set to $\mu_{ug} = 7.37$ Å and $\mu_{u0} = 0.737$ Å \cite{Kishida2000gigantic}. The probe is modeled as a Gaussian pulse with a width of $2T$, where $T = 2 \pi/ \Omega$ is the the pump period, while the pump is treated as a continuous wave acting only on the excitonic states, a valid approximation given that the experimental MIR pump duration is much longer than that of the NIR probe. The probe field peaks at 0.9 MV/cm, and the pump field strength is set to 1.8 MV/cm. We numerically integrate the Lindblad master equation using a fourth-order Runge–Kutta method with a time step of $0.001 T$, to obtain the time-dependent density matrix.

\begin{figure}[h]
\centering
\includegraphics[width=0.75 \textwidth]{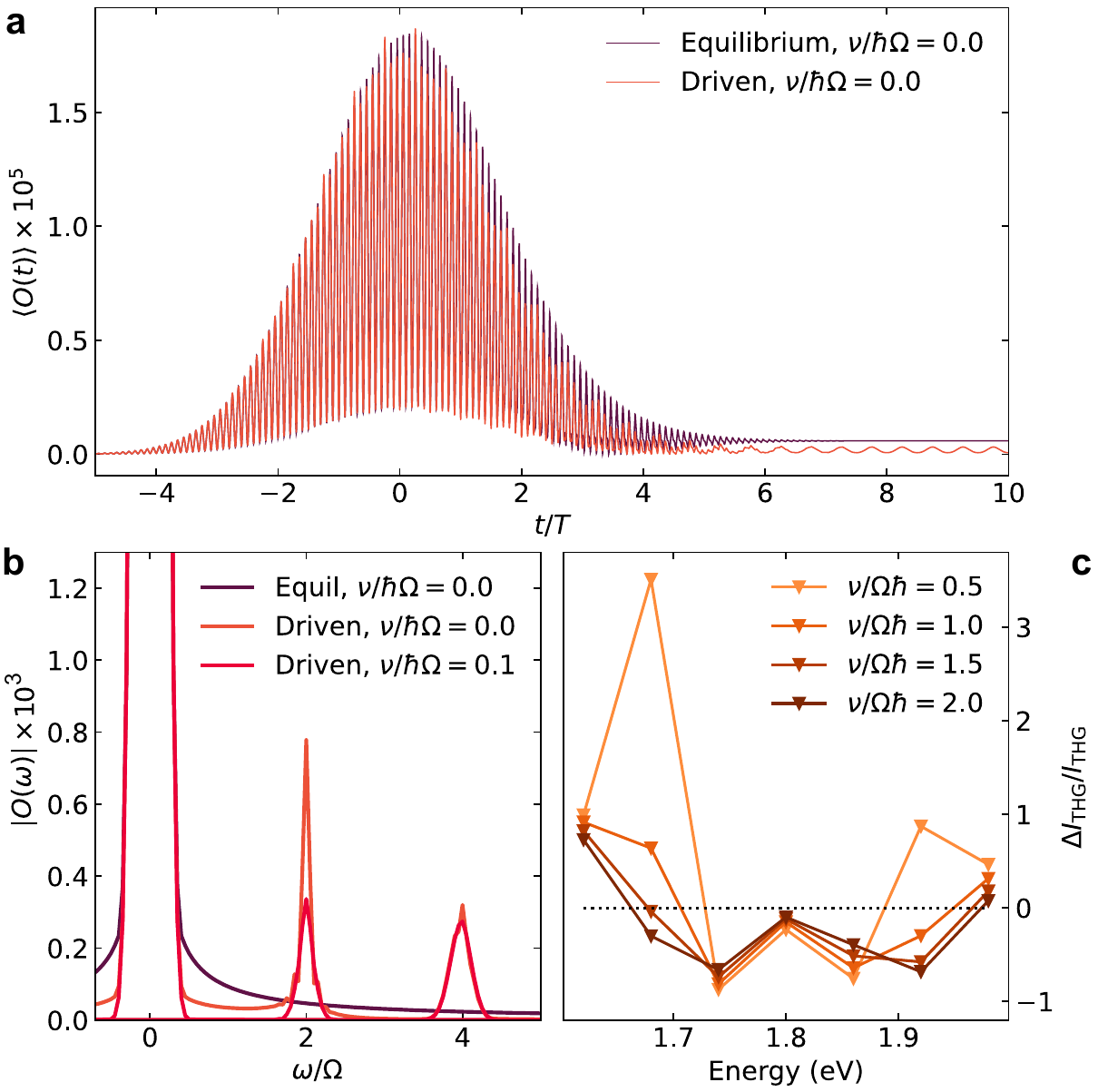}
\caption{\textbf{Effect of exciton lifetime on Floquet dressing.} \textbf{a}. Expectation value of the odd excitonic state occupation as a function of time for driven and undriven systems (without recombination effects).\textbf{b}. Fourier transform of the odd excitonic state occupation at equilibrium and under driving with and without exciton recombination \textbf{c}. Differential THG intensity change for various probe frequencies $\omega_p$ and recombination rates $\nu$.}\vspace{-7mm}
\label{fig:Coherence}
\end{figure}

We first track the occupation of the odd excitonic state, $\langle O(t) \rangle = \text{Tr}[\rho(t) O]$, where $O = \ket{u}\bra{u}$, as shown in Fig.~\ref{fig:Coherence}a. In the absence of both pump and recombination, the probe pulse alone populates the odd excitonic state, resulting in a long-lived occupation. When the system is driven, the occupation oscillates in time, reflecting the coherent rotation of the odd exciton into the even state under the pump field. This behavior is also evident in the Fourier transform of the exciton occupation shown in Fig.~\ref{fig:Coherence}b. Unlike the equilibrium case, the driven system exhibits peaks at even integer multiples of the pump frequency. This is consistent with the exact solution (Eq.~\eqref{eq:Ex_u}), where the Fourier decomposition of the dressed exciton reveals that odd-parity components oscillate at even harmonics. Introducing a finite recombination rate reduces the spectral weight of these peaks, reflecting the damping of coherent oscillations due to the finite exciton lifetime.

Next, we investigate how the exciton lifetime affects the Floquet modulation of the THG spectrum. For probe energies between 0.54 and 0.68 eV, we compute the time-dependent expectation value of the dipole operator, $\langle X(t) \rangle$, and extract its Fourier component near $3\hbar\omega_p$ for varying lifetimes. The intensity is determined by integrating the magnitude square of the spectral weight within $\pm 50$ meV of $3\hbar\omega_p$. The change in intensity between the equilibrium and driven systems is shown in Fig.~\ref{fig:Coherence}c. For all lifetimes, the main THG peak at $3\hbar\omega_p = \epsilon_{ex}$ is suppressed, although the degree of suppression decreases as the recombination rate increases. The most pronounced lifetime dependence appears at the sideband features at $3\hbar\omega_p = \epsilon_{ex} \pm \hbar\Omega$. Spectral weight emerges at these frequencies when the exciton inverse lifetime is shorter than the pump frequency, consistent with Floquet sideband formation. In contrast, when the inverse lifetime exceeds the pump frequency, this spectral weight is diminished. This reduction reflects the fact that Floquet sidebands require the exciton to persist over multiple pump cycles: if the exciton decays faster than one pump cycle, it cannot be coherently dressed by the drive. We conclude that under our experimental conditions the excitonic state is coherently rotated by the midinfrared pump and the amplitude of Floquet sidebands provides a clear indication for the degree of coherence of the driven state. \vspace{-3mm}

\section{Ruling out competition with sum-frequency generation}\label{sec:SFG_competition}\vspace{-3mm}

Owing to their strong optical nonlinearity, the Floquet-modulated third-harmonic generation (THG) of the driven Hubbard excitons is accompanied by emergent sum-frequency generation (SFG) peaks. We experimentally observe two SFG peaks at $2\hbar\omega_{\mathrm{NIR}}+\hbar\Omega_{\mathrm{MIR}}\approx 1.3$ eV and $4\hbar\omega_{\mathrm{NIR}}+\hbar\Omega_{\mathrm{MIR}}\approx 2.4$ eV, corresponding to third- and fifth-order nonlinear processes, respectively. Only odd-order nonlinearities are symmetry-allowed in this centrosymmetric compound, whereas fourth-order terms at the Floquet sideband energies $3\hbar\omega_{\mathrm{NIR}}\pm\hbar\Omega_{\mathrm{MIR}}$ are symmetry-forbidden.

\begin{figure}[h]
\centering
\includegraphics[width=\textwidth]{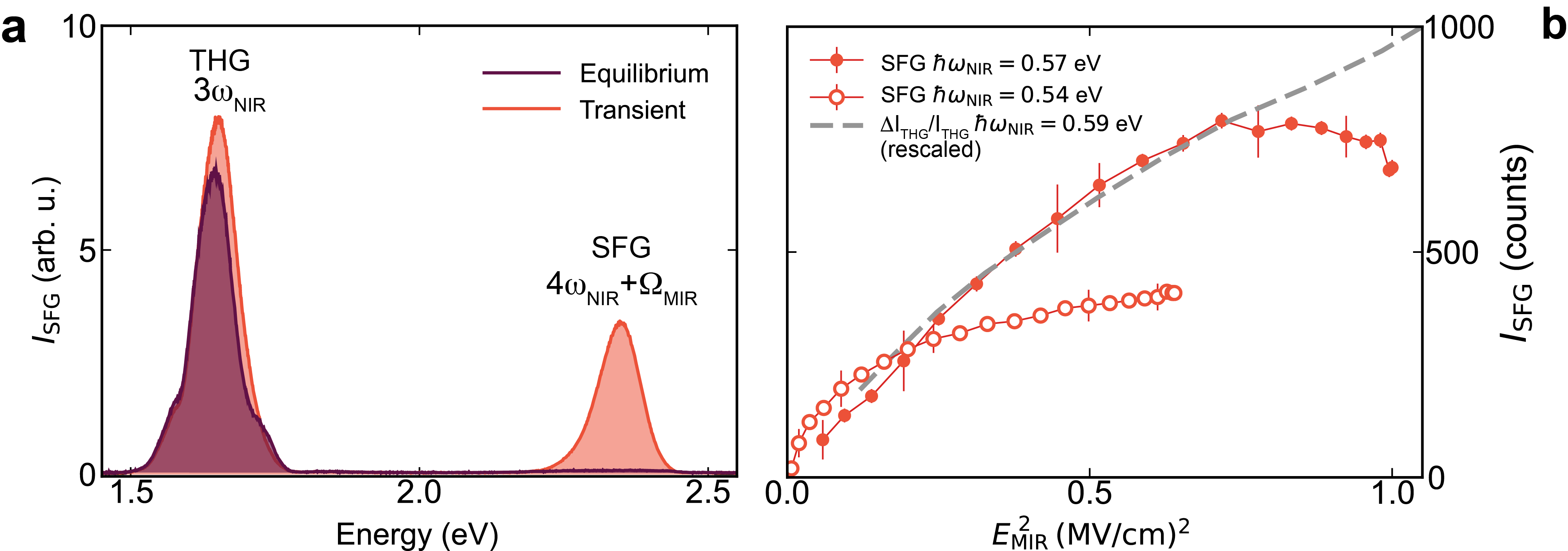}
\caption{\textbf{Fifth-order sum-frequency generation.} \textbf{a}. Third-harmonic ($3\hbar\omega_{NIR}$) and fifth-order SFG spectra ($4\hbar\omega_{NIR}+\hbar\Omega_{MIR}$) of a 0.54 eV NIR probe at equilibrium (purple) and after a 0.12 eV pump (orange, $\Delta t=0$ fs, 0.8 MV/cm field strength). \textbf{b}. Driving field dependence of the SFG process at selected probe energies and comparison with rescaled absolute value of the THG suppression.}\vspace{-7mm}
\label{fig:SFG}
\end{figure}

To rule out the possibility that the observed THG suppression arises from competition with SFG processes, we perform two independent checks: (1) we confirm that SFG peaks appear not only when THG is suppressed but also when it is enhanced at the Floquet sidebands, and (2) we measure the SFG intensity as a function of the MIR driving field (see Fig.~\ref{fig:SFG}). Both SFG peaks increase in intensity under MIR excitation, regardless of whether the corresponding THG response increases or decreases as function of the NIR drive energy (Fig.~\ref{fig:SFG}a). This observation rules out a sole photon transfer from THG to SFG as the origin of the observed modulation.

Following Ref.~\cite{shan2021giant}, we further examine the fluence dependence of the SFG peaks at $4\hbar\omega_{\mathrm{NIR}}+\hbar\Omega_{\mathrm{MIR}}$ (Fig.~\ref{fig:SFG}b). If the fluence-dependent growth of the SFG signal is slower than that of the THG peak, then a redistribution of photons from THG to SFG processes cannot fully account for the THG peak modulation. We find that the SFG intensity exhibits a saturating or nonmonotonic trend, particularly pronounced at 0.57 eV, with a slower field-dependent growth than the THG signal well below 1 MV/cm. These findings are consistent with observations in the context of SHG processes in Ref.~\cite{shan2021giant}, where the second-order nonlinearity is Floquet modulated, and indicate that the nonlinear susceptibility $\chi^{(3)}(\omega)$ is directly modulated by the MIR field.

The breakdown of quadratic scaling of the SFG intensity at these field strengths implies that Floquet effects are essential to understanding the modulation of optical nonlinearities. The SFG arises from a fifth-order process involving four probe photons and one pump photon, such that $\chi_{SFG}^{(5)} \propto E_{\mathrm{MIR}}$ and the intensity is expected to scale as $E_{\mathrm{MIR}}^2$. The observed deviation from this scaling already below 1 MV/cm indicates that higher-order processes contribute at a comparable level and cannot be neglected.

We now theoretically estimate the field strength at which the SFG and THG intensity responses begin to diverge. At low pump fluence, the differential THG intensity can be expanded in the pump field, and the modification of $\chi^{(3)}$ can be attributed to a fifth-order process involving two pump photons. However, at higher field strengths, contributions from higher-order terms lead to deviations from this $\chi^{(5)}$ behavior. In this regime, the change in $\chi^{(3)}$ reflects a coherent renormalization induced by the drive, arising from the resummation of nonlinear optical processes to all orders in the pump field, as captured by Floquet theory. In the three-state model, the peak THG intensity under midinfrared irradiation scales as $J_0 \left (\frac{\mu_{ug} \mathcal{E}}{\hbar \Omega} \right )^4$ where $\mathcal{E}$ is the MIR field, see Eq.~\eqref{eq: THG_floquet}. To leading order in the pump, the THG peak suppression goes as
\begin{equation}
\frac{\Delta I_{peak}}{I_{peak,eq}}
= 1-J_0
\left(\frac{\mu_{ug} \mathcal{E}}{\hbar \Omega}
\right)^4 
= 
\left (\frac{\mu_{ug} \mathcal{E}}{\hbar \Omega}
\right)^2 + 
\mathcal{O}(\mathcal{E}^4).
\label{eq:Bessel}
\end{equation}
We define the breakdown of this perturbative limit when higher-order corrections (those at $
\mathcal{E}^4$ or higher) account for $10 \%$ or more of the total THG suppression, that is when
\begin{equation}
\frac{|\Delta I_\mathrm{peak} /I_\mathrm{peak,eq} -  (\frac{\mu_{ug} \mathcal{E}}{\hbar \Omega}
)^2|}{\Delta I_\mathrm{peak} /I_\mathrm{peak,eq}} = 0.1
\end{equation}
The equation is satisfied when $\mu_{ug} \mathcal{E}/\hbar \Omega = 0.47$, corresponding to a suppression of $\Delta I_{\mathrm{peak}} / I_{\mathrm{peak,eq}} = 0.20$. Once the suppression of the THG peak exceeds 20\%, a complete Floquet treatment becomes necessary and the response must be understood as a pump-induced renormalization of the equilibrium third-order nonlinearity. This crossover corresponds to $E_{\mathrm{MIR}} = 0.7$–$0.8$ MV/cm in Fig. 4 of the main text, and is matched to the fields in Fig. \ref{fig:SFG} where the nonmonotonic behavior becomes apparent. \vspace{-3mm}

\section{Energy shift of the THG spectra}\label{sec:THG_shift}\vspace{-3mm}

\begin{figure}[h]
\centering
\includegraphics[width=\textwidth]{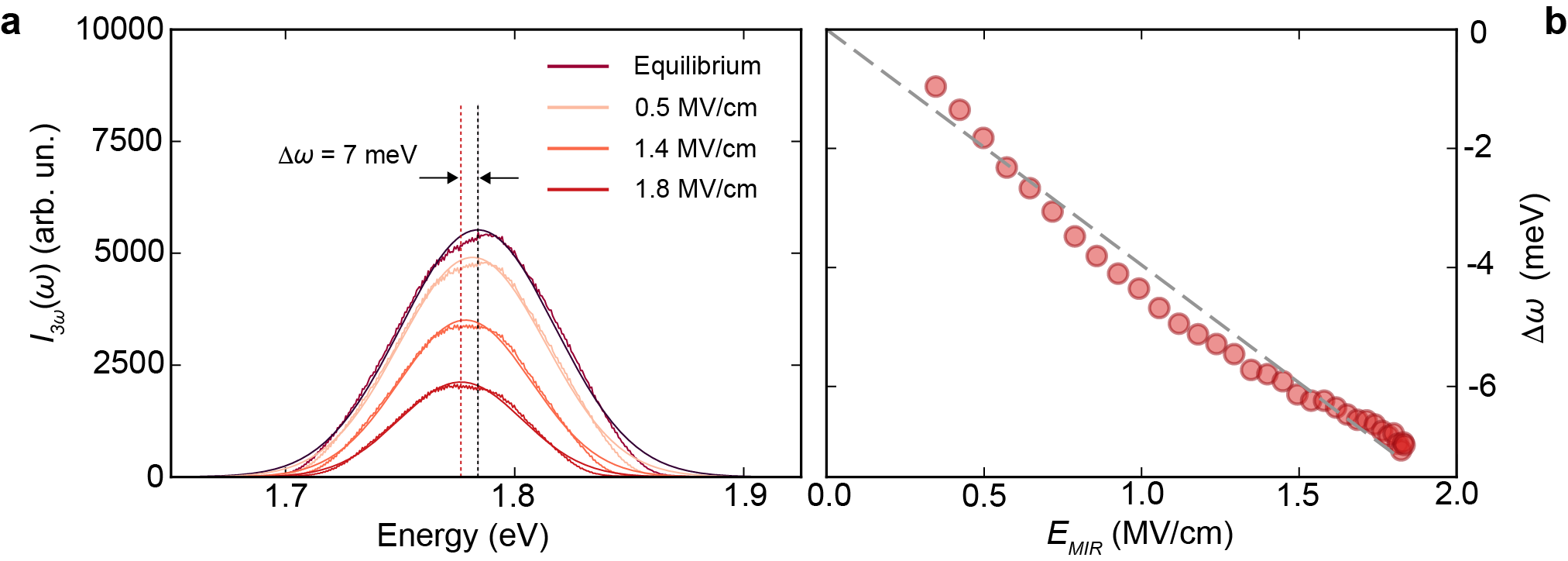}
\caption{\textbf{Fluence dependence of the THG photon energy.} \textbf{a}. Third harmonic spectra of a $0.59$~eV NIR probe at equilibrium (purple) and after a $0.12$~eV pump (orange, $\Delta t=0$~fs) for selected field strengths. Spectra are fit to Gaussian profiles (solid lines), while the thin dashed lines indicate the THG peak position at equilibrium and at 1.8 MV/cm. \textbf{b}. THG peak position extracted from Gaussian fits as in panel \textbf{a.} Error bars are within the symbol size and the dashed grey line is a guide to the eye.}\vspace{-7mm}
\label{fig:THG_peak_shift}
\end{figure}

The coherent optical dressing of quantum states by a periodic electromagnetic field induces two primary effects: state mixing and energy level shifts. These shifts arise from opposing optical Stark and Bloch-Siegert corrections \cite{Sie2017large,shan2021giant}, a hallmark of Floquet engineering in systems with non-degenerate energy levels. When the dressed states are perfectly degenerate, these energy shifts vanish, making quantum state mixing the dominant effect. In this study, the Hubbard excitons are not exactly degenerate, leading to a measurable MIR-induced shift of the THG peaks alongside the state rotation discussed in the main text. We quantify this shift by tracking the THG peak position under varying MIR field strengths, using Gaussian fits to extract peak positions (Fig. \ref{fig:THG_peak_shift}a). We keep the MIR photon energy fixed at 0.12 eV. The THG peak shifts linearly with the MIR pump field and reaches a value of 7 meV at the highest field strengths (Fig. \ref{fig:THG_peak_shift}b), consistent with a near cancellation of the Stark and Bloch-Siegert shifts and a small renormalization of the exciton energies due to coupling between the excitons and the continuum. While nonzero, this shift remains modest compared to previous reports in monolayer WS$_2$ and MnPS$_3$, where Floquet-engineered states are non-degenerate. We note that an energy shift of the excitons alone cannot explain the experimentally observed reduction in the THG intensity, as well as the emergence of sidebands at around the main $\chi^{(3)}(-3\omega; \omega, \omega, \omega)$ resonance.\vspace{-3mm}

\singlespacing
\bibliography{floquet_sr2cuo3}